\documentclass[%
 reprint,
superscriptaddress,
 amsmath,amssymb,
 aps,
floatfix,
]{revtex4-2}

\usepackage{hyperref}
\usepackage{siunitx}
\usepackage{graphicx}
\usepackage{bm}
\usepackage{dcolumn}
\usepackage{rotating}
\usepackage{multirow}
\usepackage{ragged2e}

\usepackage{tabularx}   
\usepackage{booktabs}   
\usepackage{array}      
\usepackage{makecell}   

\renewcommand{\thetable}{\arabic{table}}

\makeatletter
\def\@affil@script#1#2#3#4{%
 \@ifnum{#1=\z@}{}{%
  \begingroup
   \frontmatter@affiliationfont
   \@ifnum{\c@affil<\affil@cutoff}{}{%
    \def\@thefnmark{#1}\@makefnmark
   }%
   \ignorespaces#3%
   \@if@empty{#4}{}{\frontmatter@footnote{#4}}%
  \endgroup
 }%
}%
\makeatother

\begin{document}


\title{Depth-enhanced molecular imaging with two-photon oblique plane microscopy}

\author{Kevin Keomanee-Dizon}
\email[Corresponding author: ]{kevin@capeoptics.com}
\affiliation{%
Joseph Henry Laboratories of Physics, Princeton University, Princeton, NJ, USA\\
}%

\author{Yaakov Clenman}
\affiliation{%
Joseph Henry Laboratories of Physics, Princeton University, Princeton, NJ, USA\\
}%

\author{Alejandra Duran}
\affiliation{%
Joseph Henry Laboratories of Physics, Princeton University, Princeton, NJ, USA\\
}%

\author{Sergey Ryabichko}
\affiliation{%
Lewis--Sigler Institute for Integrative Genomics, Princeton University, Princeton, NJ, USA\\
}%

\author{Pauline Hansen}
\affiliation{%
Lewis--Sigler Institute for Integrative Genomics, Princeton University, Princeton, NJ, USA\\
}%

\author{Tohn Borjigin}
\affiliation{%
Lewis--Sigler Institute for Integrative Genomics, Princeton University, Princeton, NJ, USA\\
}%

\author{Richard H. Thornton}
\affiliation{%
Department of Molecular Biology, Princeton University, Princeton, NJ, USA\\
}%

\author{Jared E. Toettcher}
\affiliation{%
Department of Molecular Biology, Princeton University, Princeton, NJ, USA\\
}%
\affiliation{%
Omenn--Darling Bioengineering Institute, Princeton University, Princeton, NJ, USA
}%

\author{Harold M. McNamara}
\altaffiliation{Present address: Department of Molecular, Cellular, and Developmental Biology, Department of Physics, and Wu Tsai Institute, Yale University, New Haven, CT 06510, USA}
\affiliation{%
Lewis--Sigler Institute for Integrative Genomics, Princeton University, Princeton, NJ, USA\\
}%

\begin{abstract}
High-numerical-aperture (NA) oblique plane microscopy enables noninvasive fluorescence imaging of subcellular dynamics without radical sample modification.
However, performance degrades at depth in multicellular specimens as scattering and refractive-index heterogeneity raise out-of-focus background.
We report a two-photon oblique plane microscope that improves resolution at depth by combining high-NA single-objective detection with multiphoton plane illumination.
The microscope achieves $\sim\!300$~nm lateral and $\sim\!650$~nm axial resolution, with single-molecule sensitivity and $<\!60$ nm localization precision \textit{in vivo}.
Compared with two-photon point scanning, the lower illumination NA delivers order-of-magnitude lower peak intensity, enabling $>\!5\times$ faster volumetric acquisition (up to $3.25 \times 10^6$ voxels~s$^{-1}$).
In multicellular contexts, near-infrared nonlinear excitation enhances contrast throughout the illumination depth by $\sim\!2\times$ and restores volumetric resolving power by $>\!2\times$ relative to linear excitation.
We demonstrate these capabilities through molecular imaging of epithelial tissue, stem-cell-derived gastruloids, and living fruit fly embryos, including multicolor transcription-factor dynamics, optogenetic subcellular control, and single-mRNA tracking, all using standard glass-based mounting.
\end{abstract}

\maketitle



\phantomsection
\addcontentsline{toc}{section}{Introduction}
Light-sheet microscopy offers a crucial window into the intrinsic complexity of the living world, from single-molecule dynamics in cells \cite{Chen_lattice} to whole-brain neural activity in behaving animals \cite{KKD_2020,Luu_2024}.
By restricting illumination to the vicinity of the detection depth of field, it permits fast, plane-wise acquisition with high contrast, inherent optical sectioning, and low photodamage.
However, most implementations rely on multi-objective geometries and custom sample holders that are incompatible with standard glass-based mounting protocols \cite{Stelzer_2021}.
Single-objective implementations, such as oblique plane microscopy (OPM) \cite{Dunsby_2008,Bouchard_2015}, preserve standard sample mounting while enabling high-NA ($>\!1$) detection via refractive-index-mismatched remote focusing \cite{Yang_2019,Millett-Sikking,Sapoznik}.
In multicellular specimens, scattering and refractive-index heterogeneity deform and thicken one-photon (1P) light sheets.
Wide-field detection then integrates out-of-plane fluorescence, degrading  contrast, spatial resolution, and the fidelity of weak molecular signals.

Two-photon (2P) excitation helps bypass these limits: longer wavelengths scatter less, and the quadratic nonlinearity spatially concentrates fluorescence to the focal volume, suppressing background and improving sectioning at depth \cite{Denk_1990,Wu_2021,Xu_2024}.
To date, most multiphoton imaging detects signal serially, from a point \cite{Chen_2025} or a line \cite{Wu_2020}, or in parallel from multiple foci \cite{Zhang_2019} that demand proportionally higher power. Both limit speed.
For fast point or line scanning, increasing the voxel rate by $N$ shortens the dwell to $\tau' = \tau/N$, reducing photons per voxel at fixed average power at the specimen $P_\mathrm{s}$.
Since 2P signal $S \propto P_\mathrm{s}^2\tau$, maintaining the signal-to-noise ratio (SNR) requires $P_\mathrm{s}'\approx \sqrt{N}\,P_\mathrm{s}$.
In multifocal 2P, an $N$-fold speedup with $N$ simultaneous foci requires $P_\mathrm{s}'\approx NP_\mathrm{s}$ to maintain per-voxel SNR; otherwise it falls as $\sim$1/$N^2$ \cite{Supatto_2011}.
While SNR is fundamentally constrained by fluorophore photophysics and shot noise, in practice linear heating ($\propto$ average power) and higher-order nonlinear photodamage ($\propto$ peak intensity) make fast, gentle volumetric imaging at subcellular resolution challenging for common 2P methods.

Multiphoton plane illumination offers advantages in both speed and noninvasiveness \cite{Supatto_2011,Truong_2011}.
First, orthogonal illumination parallelizes excitation along the propagation direction: $N$ voxels in the plane are excited and read out simultaneously on the camera; at fixed voxel rate the per-voxel dwell equals the exposure time (not $\tau/N$ as in 
fast scanning with a point detector) \cite{Supatto_2011}.
This exposure can be orders of magnitude longer than the equivalent 
point-scanning dwell, proportionally increasing photon accumulation at the same average power.
Second, because axial resolution in light-sheet imaging is set by the sheet thickness convolved with the detection depth of field (not a tight 3D focus), the illumination NA can be reduced by $\alpha\!\approx\!5$–$10$ while preserving the point-scan axial point-spread function (PSF), which lowers the peak (instantaneous) intensity by $\alpha^2$ \cite{Truong_2011}.
Since 2P signal scales as $S \propto I_\mathrm{peak}^2 \tau$, with $I_{\text{peak}}\propto \mathrm{NA}^{2}$, this NA reduction depresses signal by $\alpha^4$.
Compensating at constant SNR, the per-voxel dwell can increase by 
$\alpha^4$ ($\sim\!6\times10^{2}$--$10^{4}$ longer) while the required peak intensity drops by $\alpha^2$ ($\sim$25--100$\times$ lower) \cite{Supatto_2011,Truong_2011}.
Lower peak intensity directly suppresses higher-order ($>$2P) nonlinear processes that dominate phototoxicity \textit{in vivo} \cite{Maioli_2020}, allowing higher tolerable average powers and thus improved SNR and volumetric imaging speed without crossing photodamage thresholds \cite{Truong_2011,Maioli_2020}.

High-resolution 2P light-sheet systems have been demonstrated \cite{Planchon_2011,Truong_2011,Gao_2012,KKD_2020}, but use multi-objective geometries that complicate sample mounting and restrict sample size.
Single-objective 2P light-sheet variants support conventional mounting but operate at low detection NA ($\sim\!\!0.34$) \cite{Kumar_2018}, preventing sub-\si{\micro\meter} resolution and sufficient photon collection for single-molecule sensitivity.

To overcome these inherent optical and sample-handling tradeoffs, we developed 2P oblique plane microscopy (2P-OPM), combining high-NA single-objective detection with multiphoton plane illumination.
Refractive-index mismatch in the remote-focus space provides high-NA detection and a conventional objective-sample interface \cite{Botcherby_2007,Yang_2019,Millett-Sikking,Sapoznik}, permitting high-resolution imaging through common glass substrates.
By delivering a near-infrared (NIR) 2P light sheet through the same objective used for collection, 2P-OPM better confines excitation to the detection depth of field and retains robust volumetric resolution at depth compared to 1P.
Our system achieves raw resolutions of $\sim\!\!300$ $\si{\nano\meter}$ laterally and $\sim\!\!650$ $\si{\nano\meter}$ axially, the highest for multiphoton OPM to date.
The resulting gains in resolution, contrast, and sectioning improve low-SNR performance and enable single-molecule sensitivity \textit{in vivo}.
We show simultaneous multicolor imaging, optogenetic control and measurement of subcellular organization, and volumetric acquisition more than fivefold faster than traditional 2P point scanning.
These capabilities are demonstrated by molecular imaging across a broad range of multicellular contexts, from gastruloids to transcriptional dynamics in living fly embryos, all prepared with standard mounting.

\section*{Results}

\subsection*{High-NA OPM with 2P excitation}
\begin{figure*}[b]
\centering
\includegraphics[width=0.90\textwidth]{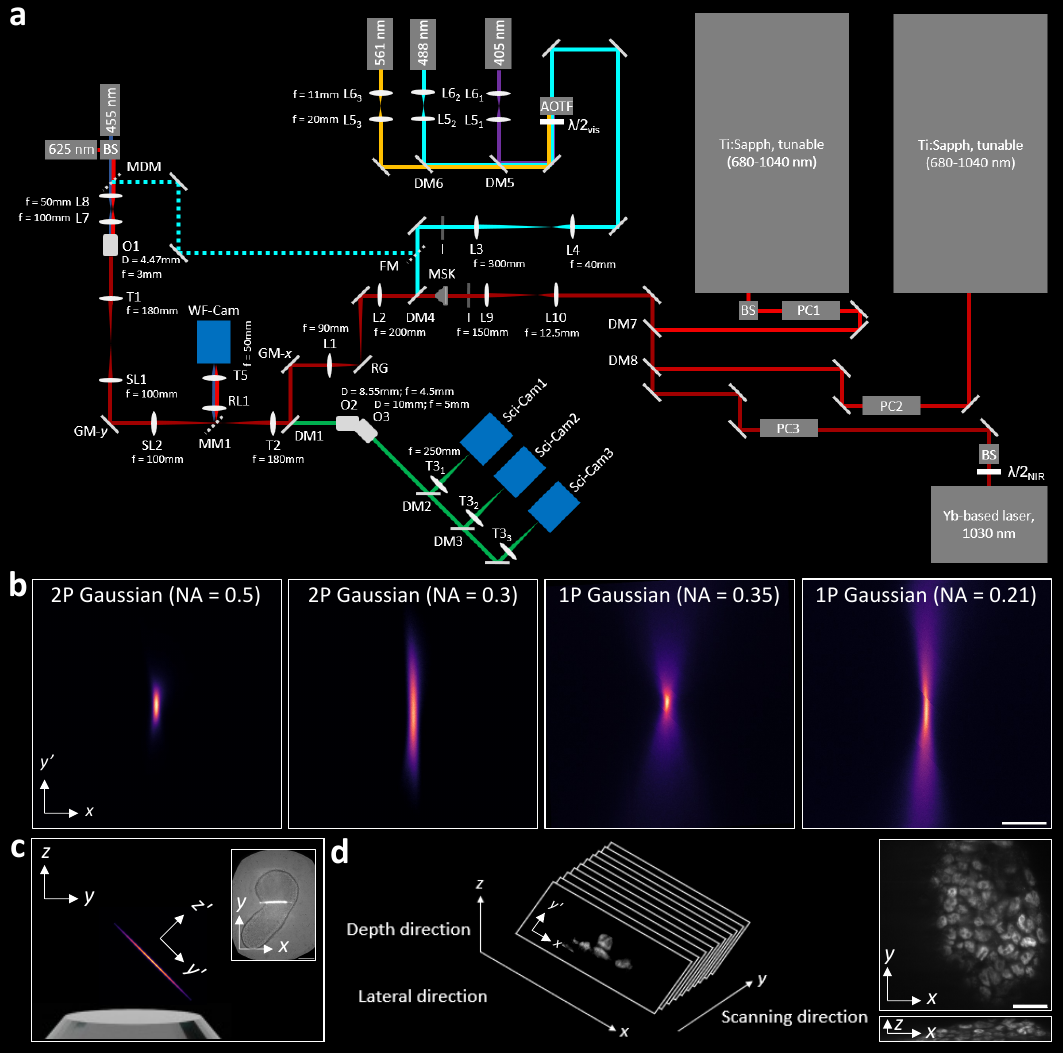}
\caption{\textbf{Two-photon oblique plane microscopy (2P-OPM). (Caption on next page)} \hfill}
\label{fig:main1}
\end{figure*}

Our design starts with high-NA OPM to enable high-resolution imaging through standard glass substrates \cite{Botcherby_2007,Yang_2019,Millett-Sikking,Sapoznik}.
Three imaging subsystems are arranged in series so that the specimen is selectively illuminated with an oblique sheet through the same objective (O1) used for collection (Fig.~\ref{fig:main1}a,c).
Fluorescence is relayed by secondary (O2) and tertiary (O3) objectives and focused onto a camera (Fig.~\ref{fig:main1}a and Supplementary Fig.~\ref{fig:2P-OPM_CAD+photo}).
This single-objective geometry provides flexibility in sample mounting similar to traditional microscopy, accommodating large samples as well as standard glass slides and plates.
High-NA OPM harnesses a deliberate refractive-index mismatch between O2 and O3 that compresses the light cone, preserving the large detection NA of O1 and yielding higher spatial resolution and photon-collection efficiency than both single-objective designs lacking this mismatch \cite{Yang_2019,Millett-Sikking,Sapoznik} and conventional multi-objective light-sheet configurations \cite{Stelzer_2021}.
In practice, light throughput and sensitivity is limited by losses in the relay and remote-focus optics.

To mitigate these losses and broaden wavelength bandwidth (for both 1P and 2P operation), we use Plössl lenses instead of thick scan lenses \cite{Voleti_2019,Negrean_2014}, both upstream and downstream of our $y$-scanning galvo mirror (GM-$y$; Fig.~\ref{fig:main1}a), used to sequentially acquire 2D images and assemble 3D volumes (Fig.~\ref{fig:main1}d) \cite{Kumar_2018,Yang_2019}.
Each Plössl lens is composed of a pair of broadband double achromats opposing each other with a small ($\sim\!\!2$ mm) air gap.
Because the scan lenses (SL1--SL2) and first set of tube lenses (T1--T2) lie in the path shared by NIR 2P excitation and visible emission, they carry broadband VIS--NIR coatings to transmit across the full band (Table~\ref{tab:parts} and Supplementary Fig.~\ref{fig:transmission}).
We also use double achromats as tube lenses T3$\textsubscript{1-3}$ with a focal length $f=250$ mm, yielding an overall system magnification of 75×.
This properly samples the diffraction-limited resolution according to the Nyquist criterion.
Three sCMOS cameras (Sci-Cam1--3) receive spectrally separated emission by a dichroic-mirror tree (Fig.~\ref{fig:main1}a), enabling simultaneous multicolor imaging:
Sci-Cam1 collects 420--474 nm; Sci-Cam2 collects 502--548 nm; Sci-Cam3 collects 579--620 nm and 662--734 nm (Supplementary Fig.~\ref{fig:spectra}).
With a 45$^\circ$ tilt relative to O1, the effective theoretical detection NA is 1.44 along the $x$ direction and 1.29 along the tilt-limited $y$ direction (Supplementary Fig.~\ref{fig:2P-OPM_geo-theory-NA}).
Transmission through our detection path (power at T3 relative to T1, excluding the dichroic-mirror tree) is $\sim\!\!60\%$ at 488 nm, an improvement of $\sim\!\!20$ percentage points over the prior laser-scanning high-NA OPM design \cite{Sapoznik}.

To extend the imaging depth of high-NA OPM, we employ 2P light-sheet excitation with ultrafast NIR laser pulses (Methods and Table~\ref{tab:parts}).
Compared to the visible light used in 1P, the longer NIR wavelengths scatter less, and the quadratic dependence of 2P-excited fluorescence signal on the excitation intensity confines fluorescence near the focus, maintaining optical sectioning even when the sheet broadens by scattering \cite{KKD_2020}.
We generate 2P light sheets by sweeping a beam focus with a pupil-conjugate $x$-galvo (GM-$x$) across the detection focal plane to form an oblique $xy'$ plane (Fig.~\ref{fig:main1}a--d); a sample-conjugate resonant galvo (RG) wobbles the sheet, suppressing shadowing artifacts from absorption or scattering by averaging over illumination angles (Supplementary Fig.~\ref{fig:2P-OPM_CAD+photo}) \cite{Huisken_2007}.
The focus is swept along the in-plane axis $x$ over 100--200 \si{\micro\meter} (chosen per sample to exclude the overilluminated edges intrinsic to scanned light sheets).
Sweeping the beam in time keeps the instantaneous intensity at the focus high at each position, while redistributing it along $x$ over the exposure; the camera integrates these passes into an effectively uniform sheet.
By contrast, a static sheet of length $x$ distributes power over all $x$ simultaneously, imposing a $1/x^{2}$ per-voxel penalty and a $1/x$ plane-integrated loss (at fixed average power), whereas a swept focus avoids the $1/x^{2}$ per-voxel penalty and keeps the plane-integrated 2P signal $\sim$constant with $x$ \cite{Truong_2011}.

Our instrument operates in both 2P and 1P modes (Fig.~\ref{fig:main1}a and Supplementary Fig.~\ref{fig:2P-OPM_CAD+photo}), enabling direct quantitative comparison on the same biological sample.
In either mode, a pupil-conjugate adjustable iris (I) sets the illumination NA and thus controls the light-sheet thickness and propagation length.
For 2P, we typically use NAs of 0.3--0.5; at $\mathrm{NA}_{\mathrm{Ill}}=0.3$, the measured axial resolution matches that of conventional 2P point scanning at NA $\approx$ 1.3 (Fig.~\ref{fig:main1}b and Supplementary Fig.~\ref{fig:2P-OPM_beams}).
For 1P, lower NAs of 0.21--0.35 produce comparable light-sheet dimensions, as the shorter visible wavelengths offset the absence of quadratic intensity narrowing.
At matched voxel rate and axial resolution, this 2P NA reduction (from $\sim$1.3 to 0.3) permits the per-voxel dwell to increase by $\alpha^{4}$, corresponding to $\sim$350$\times$ longer dwell for our NA range at fixed per-voxel SNR.
Because the peak intensity scales as $I_{\text{peak}}\propto \mathrm{NA}^{2}$, this 4.3$\times$ reduction in illumination NA lowers peak intensity by 19$\times$ at equal average power.
We evaluated propagation-invariant, Bessel-like beams \cite{ChenChakraborty_2020,ExD_KKD,Saitou_2023} and found a $\sim\!\!3\times$ longer usable sheet while maintaining a thin $\sim\!\!0.5$~\si{\micro\meter} waist (Methods; Supplementary Figs.~\ref{fig:layer-cake_sim}–\ref{fig:2P-OPM_beams}, \ref{fig:2P-OPM_Bessel}).
Despite the extended field, we favored Gaussian sheets for bioimaging. A layer-cake mask splits the beam into incoherent beamlets \cite{ChenChakraborty_2020} that stretch the focus along the propagation axis ($y'$), so the same average power lights a longer, dimmer line and yields less in-focus 2P signal per point than a Gaussian sheet, which packs its power into one bright lobe.

\newpage
\onecolumngrid

\newpage

\begin{figure*}
\justifying
\small{
\noindent
FIG. 1. {\bf Two-photon oblique plane microscopy (2P-OPM). (Continued)}\\
\textbf{(a)} Optical schematic. Two Ti:sapphire lasers and one ytterbium laser provide 2P excitation (red path); continuous-wave 405 nm, 488 nm and 561 nm lasers provide 1P excitation (blue path). The two paths merge at DM4, where a pupil-conjugate adjustable iris (I) sets the illumination NA, and a layer-cake phase mask (MSK) can optionally generate a Bessel-like focus. The beam is relayed via a resonant galvo (RG) and scanning galvos GM-$x$ and GM-$y$ through scan lenses SL1--SL2 and tube lenses T1--T2 to the rear focal plane of the primary objective O1, which launches the oblique sheet into the specimen. Fluorescence is collected by O1, relayed by remote objectives O2 and O3, split into three spectral channels by DM2, DM3 and a broadband mirror, and focused onto Sci-Cam1--3 by T3$\textsubscript{1-3}$. A flip mirror (FM) selects a transmitted-illumination mode and movable mirror MM1 diverts detected light to a wide-field inspection camera (WF-Cam). Complete details of the optical path are given in Methods; components are listed in Table~\ref{tab:parts}; see also Supplementary Fig.~\ref{fig:2P-OPM_CAD+photo}.\\
\textbf{(b)} Experimental images of fluorescence excited by 2P Gaussian beams at NA 0.5 (column 1) and NA 0.3 (column 2), and 1P Gaussian beams at NA 0.35 (column 3) and NA 0.21 (column 4), which are rapidly swept in the $x$ direction to create light sheets. See Supplementary Fig.~\ref{fig:2P-OPM_beams} for intensity profiles. Scale bar, 2 \si{\micro\meter}.\\
\textbf{(c)} A 2P light sheet is launched obliquely at $45^\circ$ from O1. The $y$, $z$, $y'$ and $z'$ directions are indicated. 2D data acquired by collecting fluorescence (light cones not shown) with the same objective. Inset: transmitted light image of a 120-hr gastruloid with the light sheet overlaid. Scale bar, 100 \si{\micro\meter}.\\
\textbf{(d)} Scanning GM-$y$ in $y$ sequentially acquires oblique 2D light-sheet images, with width $x$, length $y'$, and depth $z$ to form a 3D volume. Inset: $xy$ (top) and $xz$ (bottom) orthoslices of the 3D reconstructed gastruloid from the inset in \textbf{(c)}. See also Methods and Supplementary Fig.~\ref{fig:recon}. Scale bar, 20 \si{\micro\meter}.
}
\end{figure*}

\newpage
\twocolumngrid

A transmitted wide-field mode (WF-Cam) provides low-magnification view-finding and macroscopic global context, complementing high-resolution fluorescence imaging (Fig.~\ref{fig:main1}c,d and Supplementary Fig.~\ref{fig:WF-Cam_photo+embryo}).
It also serves as a label-free photoperturbation indicator, with bright-field imaging offering a simple readout of specimen health (Supplementary Movie~\href{https://kdizon.github.io/2P-OPM/movies/WF-embryo_movie.mp4}{1}).

\begin{figure*}[t]
\centering
\includegraphics[width=\textwidth,height=535pt,keepaspectratio]{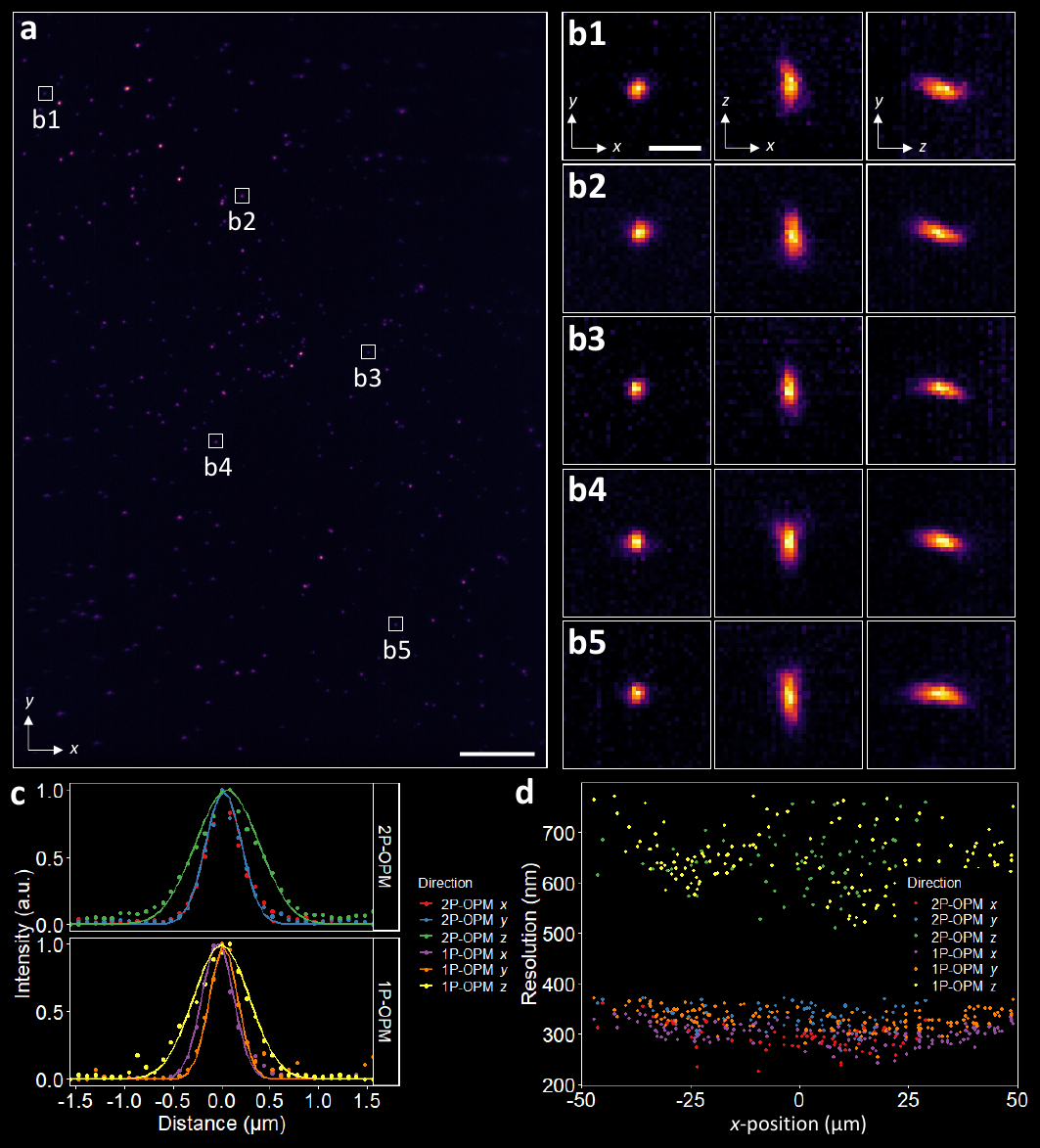}
\caption{\textbf{2P-OPM performance.}\\
\textbf{(a)} $xy$ MIP of a beads field captured with 2P-OPM.  Scale bar, 10 \si{\micro\meter}. See Supplementary Table~\ref{tab:imaging} for full imaging conditions.\\
\textbf{(b1--b5)} Zoomed-in regions from \textbf{(a)} (left) and corresponding $xz$ (middle) and $yz$ (right) MIPs, showing tight and uniform resolution throughout the volume. Scale bar, 1 \si{\micro\meter}.\\
\textbf{(c)} $x$, $y$, and $z$ line intensity profiles through representative 2P-OPM and 1P-OPM PSFs (from the bead fields in Supplementary Fig.~\ref{fig:2P+1P-OPM_PSF}), showing comparable resolution in all three dimensions. Color points: raw data; solid lines: Gaussian fit; a.u.: arbitrary units.\\
\textbf{(d)} $x$, $y$, and $z$ resolution, as measured by the FWHM, across the $x$-field of view for 2P-OPM ($N = 93$) and 1P-OPM ($N = 135$), from the bead fields in Supplementary Fig.~\ref{fig:2P+1P-OPM_PSF}. The mean $x$, $y$, and $z$ FWHM $\pm$ SD values are 2P-OPM, 292 $\pm$ 26 nm, 331 $\pm$ 21 nm, 653 $\pm$ 50 nm, respectively; and 1P-OPM, 283 $\pm$ 27 nm, 324 $\pm$ 28 nm, 613 $\pm$ 60 nm, respectively.}
\label{fig:main2}
\end{figure*}

\subsection*{Point spread function and resolution benchmarking}
We benchmarked the spatial resolution using 50 nm fluorescent beads immobilized on a coverslip (Methods; Fig.~\ref{fig:main2}a,b1--b5 and Supplementary Fig.~\ref{fig:2P+1P-OPM_PSF}a,b; Supplementary Table~\ref{tab:imaging}).
For each modality, $>90$ isolated beads were analyzed from $xy$, $xz$, and $yz$ maximum intensity projections (MIPs) via 1D Gaussian fits to line profiles.
For 2P-OPM ($\lambda_{\mathrm{Ill}}=925$~nm; NA$_{\mathrm{Ill}}$ = 0.3), the mean full width at half maximum (FWHM) $\pm$ SD is $x = 292 \pm 26$~nm, $y = 331 \pm 21$~nm, and $z = 653 \pm 50$~nm;
for 1P-OPM ($\lambda_{\mathrm{Ill}}=488$~nm; NA$_{\mathrm{Ill}}$ = 0.21), the corresponding values are $x = 283 \pm 27$~nm, $y = 324 \pm 28$~nm, and $z = 613 \pm 60$~nm (Fig.~\ref{fig:main2}c,d).
Because the beads sample is homogeneous and weakly scattering, and because the detection optics and illumination sheet thickness were matched between 2P and 1P (Fig.~\ref{fig:main1}a,b and Supplementary Fig.~\ref{fig:2P-OPM_beams}), the measured PSFs for 2P- and 1P-OPM have similar dimensions along $x$, $y$, and $z$.

Using off-the-shelf achromatic doublets for T3\textsubscript{1–3}, rather than specialized tube lenses that often trade spectral bandwidth for field, delivers uniform resolution across a $100 \times 100 \times 15$ ($xyz$) \si{\micro\meter}$^{3}$ volume (Fig.~\ref{fig:main2}b1--b5,d), a lateral span consistent with early high-NA OPM designs~\cite{Yang_2019}.
The lateral PSF is mildly anisotropic ($x<y$; $y/x \approx 1.13$), characteristic of high-NA OPM due to the O3 tilt, and matches our design calculations and theoretical simulations (Methods; Supplementary Figs.~\ref{fig:2P-OPM_geo-theory-NA} and~\ref{fig:2P-OPM_sim}).
Lateral resolution is set by the detection NA and emission wavelength 
($\lambda_{\mathrm{Det}}/2\mathrm{NA}_{\mathrm{Det}}$) and is relatively insensitive 
to the illumination conditions, while axial resolution is set by light-sheet thickness 
($\lambda_{\mathrm{Ill}}/2\mathrm{NA}_{\mathrm{Ill}}$) convolved with the detection depth of field ($\lambda_{\mathrm{Det}}/n(1-\cos\theta_{\mathrm{Det}})$) \cite{Gao_2014}.
As expected, a $\sim\!\!0.9$-\si{\micro\meter}-thick illumination plane coincident with and thinner than the detection depth of field improves axial resolution by over 30\% compared to wide-field illumination.
In contrast to earlier 2P single-objective light-sheet designs limited to low detection NA \cite{Kumar_2018}, the index-mismatched remote focus preserves the high effective NA of O1 (up to 1.44; Supplementary Fig.~\ref{fig:2P-OPM_geo-theory-NA}); these measurements thus represent the highest spatial resolving power for 2P single-objective light-sheet microscopy.

\begin{figure*}[t]
\centering
\includegraphics[width=\textwidth,height=451pt,keepaspectratio]{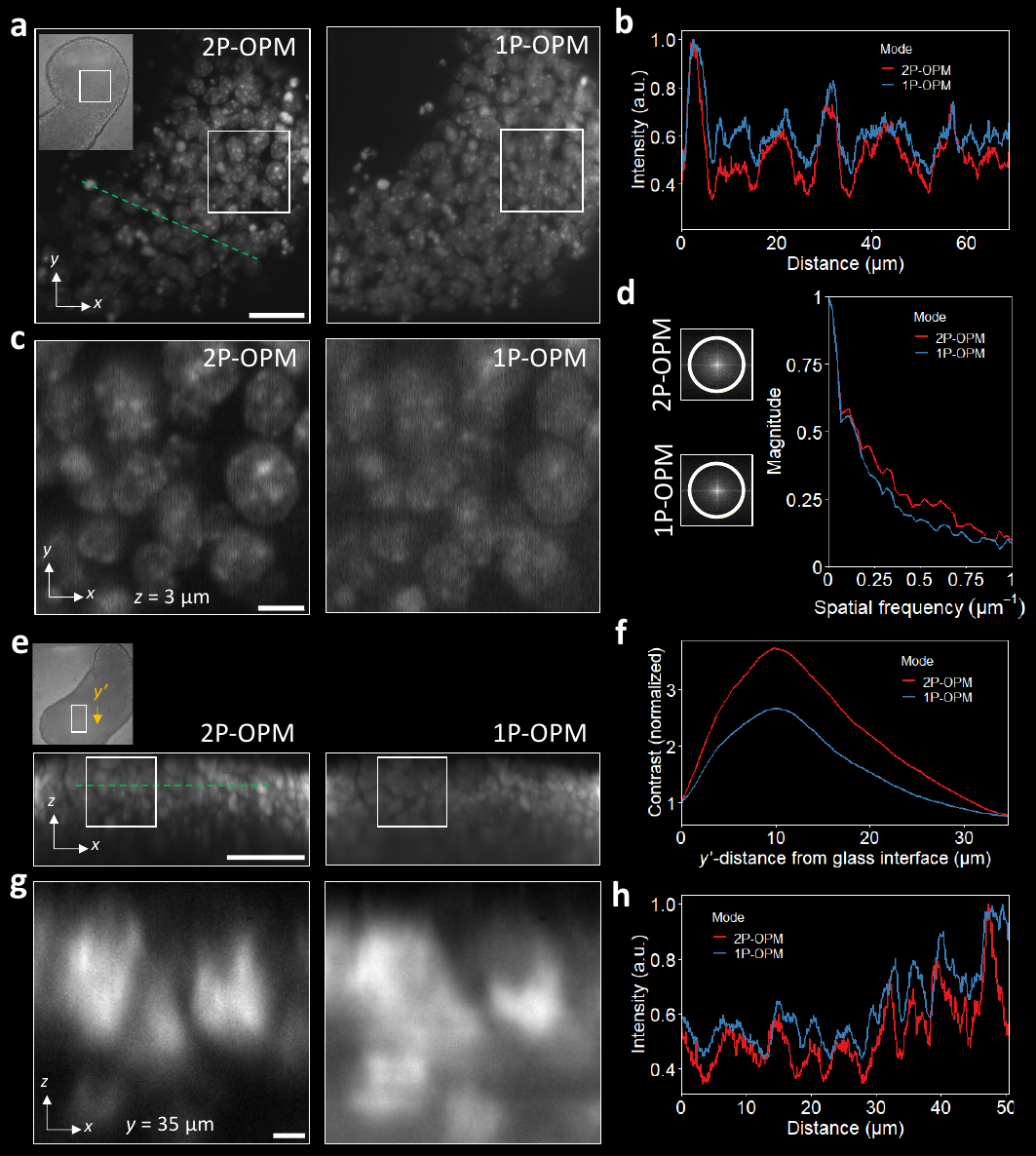}
\caption{\textbf{2P-OPM improves the contrast and resolution in volumetric imaging of multicellular systems.}\\
\textbf{(a)} $xy$ MIPs of a 120-hr DAPI-stained gastruloid, captured with 2P-OPM (left) and 1P-OPM (right). Inset: transmitted light image showing the $100 \times 115 \times 15$ \si{\micro\meter}$\textsuperscript{3}$ volume imaged. See also Supplementary Movie~\href{https://kdizon.github.io/2P-OPM/movies/oids_movie.mp4}{2}. Scale bar, 20 \si{\micro\meter}.\\
\textbf{(b)} Intensity profiles along the dashed green line in \textbf{(a)}, showing that 2P-OPM provides higher SNR than 1P-OPM. a.u.: arbitrary units.\\
\textbf{(c)} Zoomed-in $xy$ slice from the box in \textbf{(a)}, highlighting cleaner images with 2P-OPM. Scale bar, 5 \si{\micro\meter}.\\
\textbf{(d)} Fourier transforms of $xy$ MIPs in \textbf{(a)}. Resolution bands (white circles at 1 \si{\micro\meter}$\textsuperscript{-1}$) highlight the increased spatial frequency content  of 2P-OPM compared to 1P-OPM. The average amplitudes of the Fourier spectra (right) show that 2P-OPM reaches the noise floor more slowly, indicating superior effective resolution.\\
\textbf{(e)} $xz$ MIPs of a 120-hr gastruloid stained by immunofluorescence for FOXC1, recorded with 2P-OPM (left) and 1P-OPM (right). Top inset: transmitted light image showing the $71 \times 130 \times 30$ \si{\micro\meter}$\textsuperscript{3}$ volume imaged. Yellow arrow indicates the light-sheet propagation direction  ($y'$), corresponding to the $y'$-distance from the glass interface. Scale bar, 20 \si{\micro\meter}.\\
\textbf{(f)} Quantification of image contrast as a function of $y'$-distance from the glass interface, showing improved contrast over the full illumination depth with 2P-OPM over 1P-OPM. Each $xz$' slice is normalized to the surface value.\\
\textbf{(g)} Zoomed-in $xz$ orthoslice from the box in \textbf{(e)}, revealing that 2P-OPM reduces background and better resolves individual cell nuclei deep in the gastruloid tissue than 1P-OPM. Scale bar, 2 \si{\micro\meter}.\\
\textbf{(h)} Intensity profiles along the dashed green line in \textbf{(e)}, demonstrating the improved SNR and optical sectioning of 2P-OPM. a.u.: arbitrary units.}
\label{fig:main3}
\end{figure*}

\subsection*{Enhanced performance in multicellular systems}
Dense multicellular aggregates (such as organoids and gastruloids) present a stringent test for optical sectioning, where scattering and out-of-focus fluorescence rapidly degrade contrast under 1P excitation.
The tight, planar confinement of 2P excitation mitigates these effects, reducing scattering-induced background.
In line with this, 2P-OPM images of gastruloids exhibit clearer subcellular structure than 1P-OPM across matched fields of view (Supplementary Movie~\href{https://kdizon.github.io/2P-OPM/movies/oids_movie.mp4}{2} and Fig.~\ref{fig:main3}a).

Quantitative analyses reinforce this improvement. Line profiles drawn across identical trajectories (Fig.~\ref{fig:main3}b) show higher SNR with 2P-OPM, reflecting the recovery of signal otherwise lost to scattering.
Zoomed $xy$ views (Fig.~\ref{fig:main3}c) resolve individual cell nuclei with greater local contrast.
Fourier spectra of the $xy$ projections (Fig.~\ref{fig:main3}d) show elevated mid- to high-spatial frequency content with 2P-OPM,
demonstrating more faithful preservation of nuclear to subnuclear structures in the images.
Together with bead-based PSF measurements in homogeneous media, these results emphasize that while detection PSFs are comparable, nonlinear excitation substantially enhances effective resolution in scattering tissue.

Axial projections further highlight this depth advantage.
In matched $xz$ views (Fig.~\ref{fig:main3}e,g), 2P excitation suppresses out-of-focus background and maintains thin optical sections 30 \si{\micro\meter} into the specimen, reached with a lower-NA, longer light sheet (Supplementary Table~\ref{tab:imaging}), whereas 1P-OPM accumulates extraneous fluorescence.
Intensity profiles taken along the dashed lines (Fig.~\ref{fig:main3}h) exhibit narrower peaks and increased SNR, consistent with confinement of excitation to the detection focal plane.
A depth-dependent contrast metric (Methods, Eq.~\eqref{eq:contrast}; Fig.~\ref{fig:main3}f), normalized at the surface, remains higher for 2P-OPM across the full 35 \si{\micro\meter} illumination depth, confirming stronger sectioning in thick, densely labeled specimens.
Averaged over this range, contrast is $>\!\!1/3$ higher for 2P-OPM than 1P-OPM.
Integrated over the same range (area above unity), the contrast is $\sim\!2\times$ larger for 2P-OPM, consistent with prior multi-objective 2P light-sheet systems \cite{Truong_2011}.

\begin{figure}[!t]
\centering
\includegraphics[width=\columnwidth]{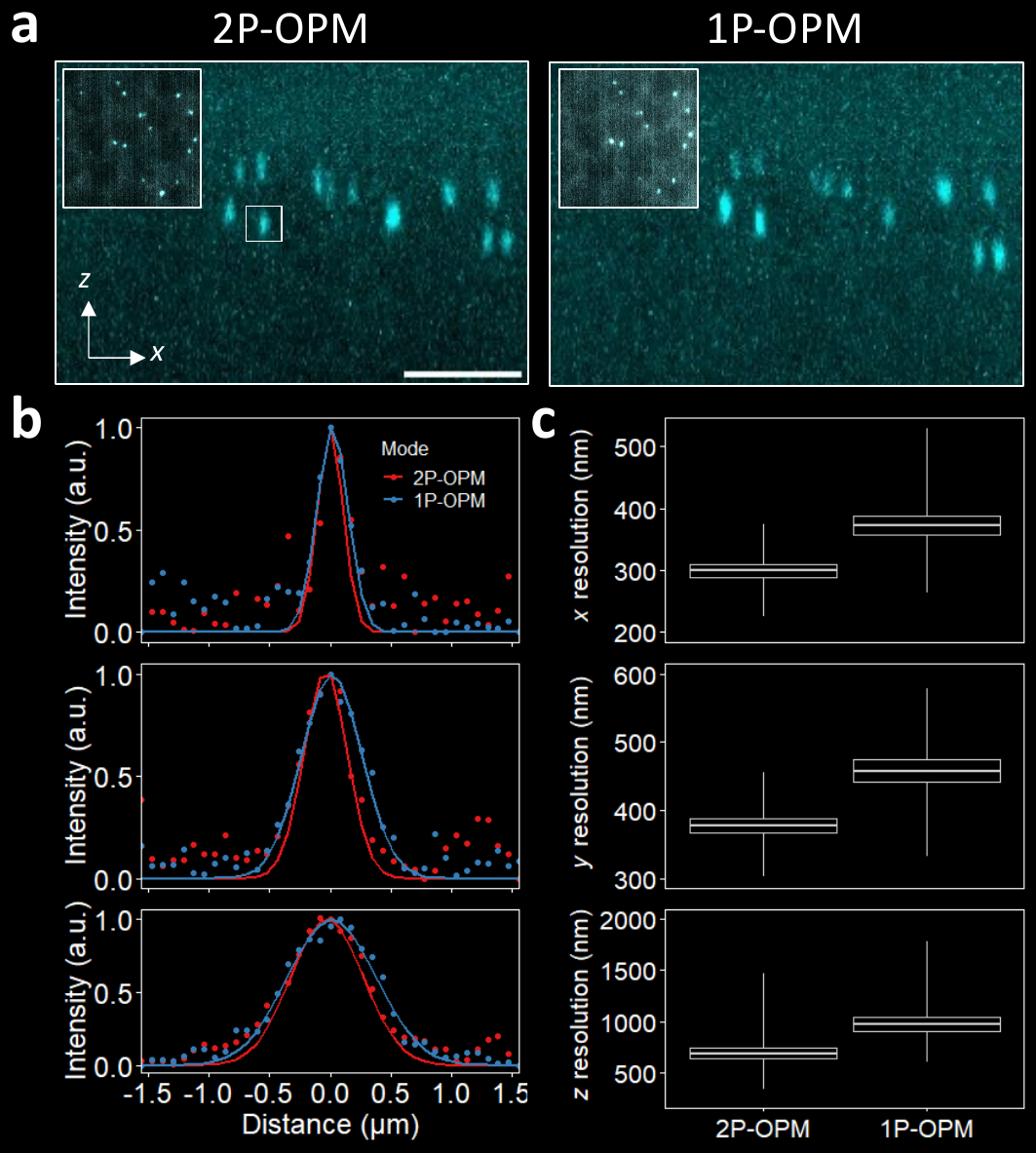}
\caption{\textbf{2P-OPM enhances molecular imaging performance in low SNR fluorescence.}\\
\textbf{(a)} $xz$ MIPs of mRNA activity for the \textit{eve} gene, labeled via smFISH, in a fixed \textit{Drosophila} embryo at NC 14, comparing 2P-OPM (left) and 1P-OPM (right). Inset: $xy$ view, highlighting activity in individual nuclei. Scale bar, 5 \si{\micro\meter}.\\
\textbf{(b)} $x$ (top), $y$ (middle), and $z$ (bottom) line intensity profiles through the transcript from the subregion indicated in \textbf{(a)}, showing better resolution in all dimensions. Color points: raw data; solid lines: Gaussian fit; a.u.: arbitrary units.\\
\textbf{(c)} $x$ (top), $y$ (middle), and $z$ (bottom) resolution, measured by the FWHM; for each modality, the same $N = 20$ individual \textit{eve} transcripts were measured in the embryo shown in \textbf{(a)}. Box bounds, mean $\pm$ standard error (not percentiles); center value, mean; whiskers, minimum to maximum. The mean $x$, $y$, and $z$ FWHM $\pm$ SD values are 2P-OPM, 299 $\pm$ 46 nm, 377 $\pm$ 50 nm, 692 $\pm$ 237 nm, respectively; and 1P-OPM, 373 $\pm$ 69 nm, 458 $\pm$ 75 nm, 975 $\pm$ 296 nm, respectively.}
\label{fig:main4}
\end{figure}

\subsection*{Molecular imaging under low-SNR conditions}
Molecular fluorescence imaging in intact tissue is often limited by background rather than diffraction.
Under linear (1P) excitation, out-of-focus fluorescence creates a near-uniform haze that raises the noise floor; the faintest signals then sit only slightly above this noise and, in practice, set the image SNR.
As the background increasingly overwhelms the signal, high-fidelity photon counts decrease and shot noise hampers the effective resolution.
In contrast, 2P-OPM spatially localizes excitation to the illumination and detection focal planes, keeping noise low so that dim signals remain resolvable.

In single-molecule fluorescence in situ hybridization (smFISH) measurements of \textit{eve} transcripts in fixed \textit{Drosophila} embryos, 2P-OPM yields cleaner, less anisotropic spots than 1P-OPM in identical fields (Fig.~\ref{fig:main4}a).
Line profiles through isolated subdiffractive puncta (Fig.~\ref{fig:main4}b) show narrower Gaussian fits along all axes.
Per-punctum SNR (Methods) was higher under 2P-OPM for all $N = 20$ matched puncta, by $\sim$1.5-fold on average (Supplementary Fig.~\ref{fig:snr}), with signal only $\sim$8\% above the local background in both modalities, placing these transcripts in the background-limited regime.
The mean spot FWHM $\pm$ SD improves from $373 \pm 69$~nm to $299 \pm 46$~nm in $x$, from $458 \pm 75$~nm to $377 \pm 50$~nm in $y$, and from $975 \pm 296$~nm to $692 \pm 237$~nm in $z$ (Fig.~\ref{fig:main4}c), reductions of $\approx 20\%$, $\approx 18\%$, and $\approx 29\%$, respectively, computed over the same $N=20$ transcripts.
These per-axis reductions yield a $\sim\!\!2.13\times$ recovery in 3D resolution (diffraction-limited puncta sharpened by $\sim\!\!53\%$).
Because the detection PSF and light-sheet thickness are matched between the modalities, these gains reflect enhanced thin-plane illumination and background rejection via 2P, with the largest benefit along $z$ where out-of-plane blur dominates.

\begin{figure*}[t]
\centering
\includegraphics[width=\textwidth,height=419pt,keepaspectratio]{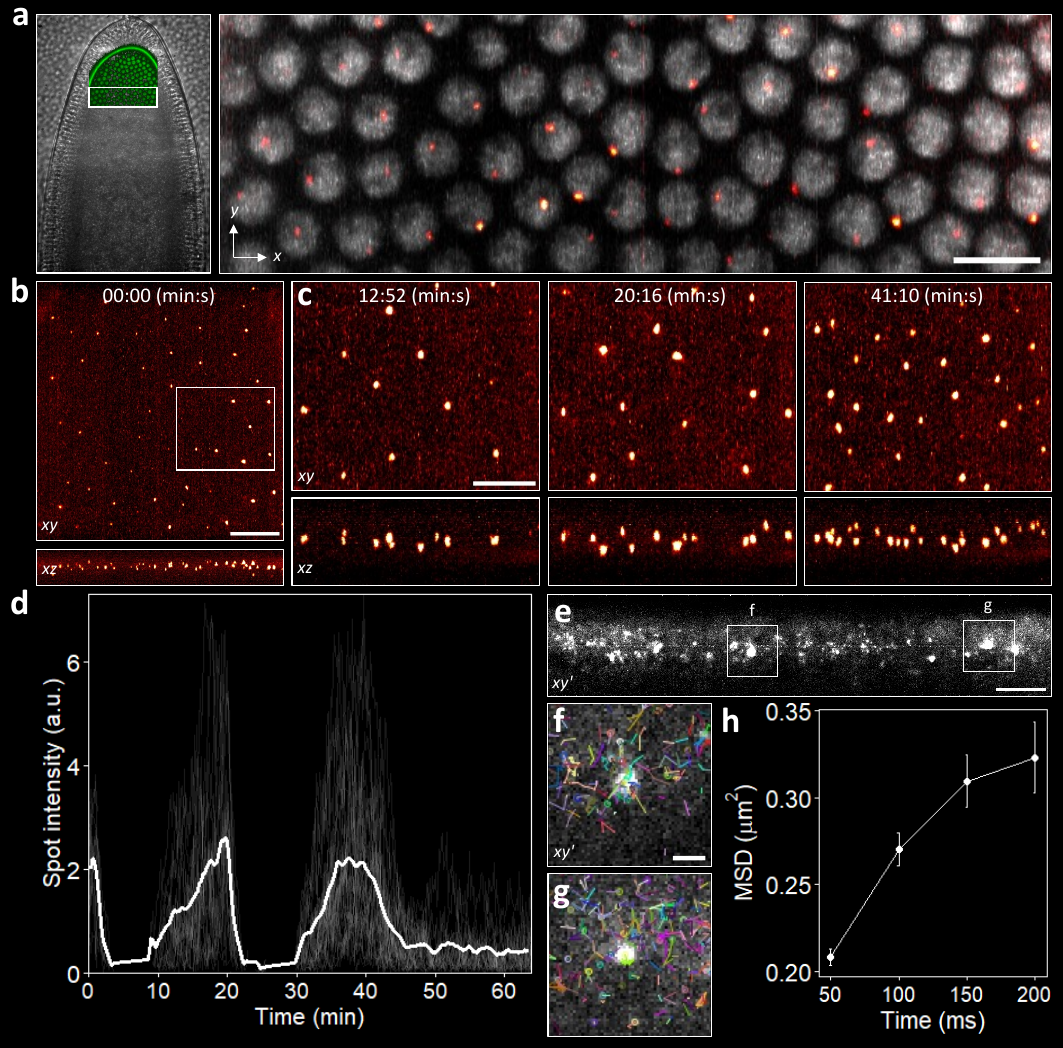}
\caption{\textbf{2P-OPM enables fast, multicolor, single-molecule imaging in living animals.}\\
\textbf{(a)} Left: Transmitted light (brightfield) image of a live fruit fly embryo at NC 14, expressing Bcd-eGFP (overlaid in green), marking the $100 \times 115 \times 15$ (\textit{xyz}) \si{\micro\meter}$^{3}$ volume imaged with 2P-OPM. The white rectangle indicates the 3.5-\si{\micro\meter}-thick slab shown at right. Right: Zoomed $xy$ MIP of Bcd-eGFP (gray) and nascent \textit{hb} transcripts, tagged with MS2-MCP-mCherry (orange hotspots). Scale bar, 10 \si{\micro\meter}.\\
\textbf{(b)} $xy$ (top) and $xz$ (bottom) MIPs from a $100 \times 105 \times 15$ \si{\micro\meter}$^{3}$ volume in a developing fruit fly embryo, showing active \textit{hb} transcription loci labeled with MS2-MCP-mNeonGreen at the end of NC 12. See also Supplementary Movie~\href{https://kdizon.github.io/2P-OPM/movies/4D-transcription_movie.mp4}{3}. Scale bar, 20 \si{\micro\meter}.\\
\textbf{(c)} Zoomed time-lapse of NC events from \textbf{(b)}. Left: beginning of NC 13; middle: peak of NC 13; right: peak of NC 14. Scale bar, 10 \si{\micro\meter}.\\
\textbf{(d)} Transcription traces of 148 \textit{hb} spots sampled every $\sim\!\!10$ s from a $100 \times 15 \times 15$ \si{\micro\meter}$^{3}$ subvolume within a $100 \times 115 \times 15$ \si{\micro\meter}$^{3}$ imaged field of a fly embryo. This subvolume corresponds to mid-anterior along the AP axis, centered at $x/L$ $\approx$ 0.45 (spanning $\sim$0.37-0.53). Mean activity is overlaid (white). a.u.: arbitrary units.\\
\textbf{(e)} Temporal standard-deviation projection showing \textit{hb} mRNAs as diffraction-limited puncta and transcription sites as larger, brighter hotspots. $100 \times 20$ \si{\micro\meter}$^{2}$ field of view captured at 20 Hz over 50 time points (Supplementary Movie~\href{https://kdizon.github.io/2P-OPM/movies/SM_movie.mp4}{4}). Scale bar, 10 \si{\micro\meter}.\\
\textbf{(f,g)} Zoomed views of the boxed regions from \textbf{(e)}, showing the first time point with single-particle trajectories overlaid from the full sequence (tracks $\ge$2 steps; colors indicate track identity). Scale bar, 2 \si{\micro\meter}.\\
\textbf{(h)} Mean squared displacement (MSD) of \textit{hb} mRNA molecules as a function of time ($\Delta t=50$--$200$~ms). Points are mean~$\pm$~standard error across $N = 3794$, 2357, 1673 and 1272 displacement pairs at $\Delta t = 50$, 100, 150 and 200~ms; the localization-corrected MSD at $\Delta t = 50$~ms yields an apparent ensemble diffusion coefficient of $D\approx1.04~\si{\micro\meter^{2}\per\second}$.}
\label{fig:main5}
\end{figure*}

\subsection*{Molecular imaging \textit{in vivo}}
Having validated improved contrast and resolution in \textit{in vitro} cell aggregates and intact fixed tissue, we next asked whether these gains translate \textit{in vivo}.
In nuclear cycle (NC) 14 embryos co-expressing Bicoid transcription factors (Bcd–eGFP) and nascent \textit{hunchback} (\textit{hb}) transcripts (MS2–MCP-mCherry), 2P-OPM enables simultaneous two-color imaging of Bcd and active transcription sites (Fig.~\ref{fig:main5}a).
Bcd shows nuclear-to-nuclear variability and subnuclear heterogeneity, with
localized enrichments evident in nearly all nuclei, mirroring recent reports of heterogeneous Bcd distributions and transient local concentration near active transcription loci \cite{Mir_2018,Munshi_2025}.

Volumetric time-lapse imaging of transcriptional activity over developmental time (NC~12–14) reveals bright \textit{hb} hotspots with high contrast in $xy$, as well as tight axial hotspots in the corresponding $xz$ projections, representing strong optical sectioning under 2P excitation (Fig.~\ref{fig:main5}b).
4D dynamics and zoomed subregions illustrate the onset and peaks of individual active transcription sites across successive nuclear cycles (Supplementary Movie~\href{https://kdizon.github.io/2P-OPM/movies/4D-transcription_movie.mp4}{3} and Fig.~\ref{fig:main5}c).

Our implementation of 2P light-sheet excitation delivers over an order-of-magnitude lower peak intensity at a given average power ($\sim\!\!4.3\times$ smaller illumination NA relative to point scanning at NA $\approx$ 1.3 yields a peak intensity of $\sim$0.5~TW~cm$^{-2}$ versus $\sim$9~TW~cm$^{-2}$, a $\sim\!\!19\times$ reduction), reducing the propensity for nonlinear photodamage.
This headroom permits higher tolerable average power and thus higher resolution sampling in time and space.
Exploiting this low-peak, high-average-power photon budget, we increased the voxel rate to $3.25\times10^{6}\,\mathrm{s^{-1}}$, over $5\times$ the $6.22\times10^{5}\,\mathrm{s^{-1}}$ used in previous work by 2P point-scanning microscopy \cite{Chen_2025}.
From a mid-anterior subvolume we then extracted 148 \textit{hb} transcription-site trajectories across NC~12–14 (Methods; Fig.~\ref{fig:main5}d);
despite $>\!1$~hr of continuous illumination with 80 mW of average power ($4\times$ higher than 2P point scanning), the fluorescence traces are not compromised, and both single-spot and mean trajectories recapitulate the expected bursty, NC dynamics observed in live embryos \cite{Garcia_2013,Chen_2025}.
The embryo showed no phenotypic signs of phototoxicity and progressed through normal sequences of development, as confirmed by a 16-hr bright-field time-lapse acquired after 2P-OPM imaging (Supplementary Fig.~\ref{fig:WF-embryo_post2p}).

Combining 2P-OPM with the MS2 stem-loop-based labeling system also allowed us to directly observe single transcripts in a living fly embryo.
MS2 provides a locus-tied burst of photons from each nascent transcript, maximizing per-molecule signal, while selective-plane 2P excitation suppresses out-of-focus background and photobleaching that would otherwise contaminate single-molecule SNR and reduce localization precision under 1P excitation in a highly scattering specimen.

With a $\sim\!1.5\times$ increase in average power (122~mW versus\ 80~mW used in Fig.~\ref{fig:main5}b--d and Supplementary Movie~\href{https://kdizon.github.io/2P-OPM/movies/4D-transcription_movie.mp4}{3}) and the 2P light sheet parked at a fixed oblique plane, the SNR is sufficient to resolve individual \textit{hb} mRNAs as they are released from the transcription site and to track their diffusion across a 100 × 20  \si{\micro\meter}$\textsuperscript{2}$ field of view at 20 Hz (Fig.~\ref{fig:main5}e and Supplementary Movie~\href{https://kdizon.github.io/2P-OPM/movies/SM_movie.mp4}{4}).
We tracked $\sim\!1.4 \times 10^3$ \textit{hb} molecules, comprising $8.5 \times 10^3$ localizations (Fig.~\ref{fig:main5}f,g).
The mean detected emission was $\sim\!1.1 \times 10^2$ photons per localization, with a mean precision of $\sigma_{\mathit{x}} = 32.9$ nm and $\sigma_{\mathit{y}'} = 58.4$ nm (Cramer-Rao lower bound; see Methods).
Single-particle tracks place the apparent ensemble diffusion coefficient of \textit{hb} transcripts at $D\approx 1.04~\si{\micro\meter^{2}\per\second}$ (Fig.~\ref{fig:main5}h), between the fast diffusion coefficient of \textit{Sox2} transcription factors in mammalian cells ($\sim\!0.89~\si{\micro\meter^{2}\per\second}$) and \textit{Zelda} transcription factors in fly embryos ($\sim\!1.55~\si{\micro\meter^{2}\per\second}$), both measured by lattice light-sheet microscopy~\cite{Chen_lattice,Mir_2018}.
This intermediate value for \textit{hb} is consistent with mRNAs diffusing more slowly than freely diffusing, non-DNA-bound transcription factors in mammalian nuclei \cite{ShavTal_2004}, and likely reflects transient interactions within the nuclear environment that slow, but do not arrest, diffusion.

\begin{figure}[!t]
\centering
\includegraphics[width=\columnwidth]{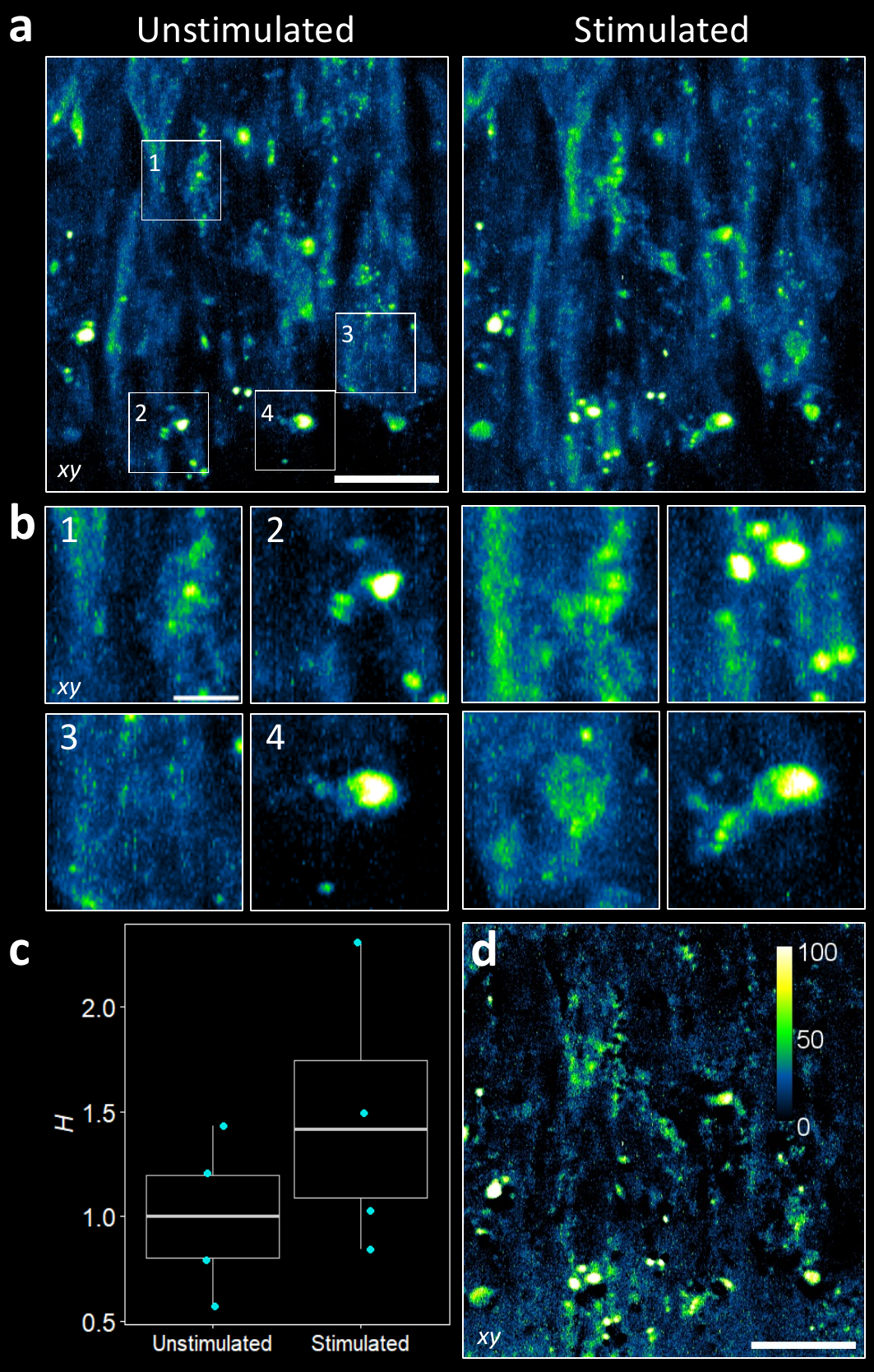}
\caption{\textbf{2P-OPM reveals light-controlled subcellular organization in living cells and tissues.}\\
\textbf{(a)} $xy$ MIPs of OptoEGFR-expressing tissues before (unstimulated) and after (stimulated) LED illumination. White boxes (1--4) highlight regions where light triggers clustering and subcellular reorganization. Scale bar, 20 \si{\micro\meter}.\\
\textbf{(b)} Zoomed views from \textbf{(a)}, showing the emergence of new puncta or brightening of pre-existing ones. Scale bar, 5 \si{\micro\meter}.\\
\textbf{(c)} Heterogeneity index $H$ measured in the ROIs from \textbf{(b)}.
Light stimulation increases $H$ by $\sim\!45\%$ across all four ROI pairs. Box bounds, mean $\pm$ standard error (not percentiles); center value, mean; whiskers, minimum to maximum. Individual values for the $N = 4$ ROI pairs defined in \textbf{(b)} are overlaid.\\
\textbf{(d)} Difference (stimulated $-$ unstimulated) image highlighting gains after stimulation. Color map localizes the enrichment of bright puncta consistent with light-induced molecular clustering. Display clipped at zero; negative values reflect residual differences between acquisitions. Scale bar, 20 \si{\micro\meter}.}
\label{fig:main6}
\end{figure}

\subsection*{Optogenetic control and molecular readout}
Optogenetics enables precise control of protein activity with light, making it a powerful tool for understanding the dynamics of cellular systems.
By combining optogenetic perturbation with high-resolution imaging, one can both trigger specific subcellular events and directly visualize their spatiotemporal responses.
The broad spectral excitation window of our hardware offers optimal pairing of imaging and optogenetic wavelengths.
2P activation of CRY2-based optogenetic constructs has been reported near $\sim\!\!850$ nm and $\sim\!\!950$ nm \cite{Taslimi_2014,Izquierdo_2018}, so we used 1030 nm for imaging to avoid inadvertent photoactivation.
This wavelength frees the visible spectrum for optogenetic control while delivering high-contrast molecular imaging throughout the tissue, with no light-induced clustering under 1030~nm illumination alone (Supplementary Fig.~\ref{fig:opto-control}).

Leveraging this spectral separation, we activated light-sensitive optogenetic epidermal growth factor receptor (OptoEGFR) with brief, wide-field blue illumination ($\lambda_{\mathrm{Stim}}=455$~nm) in epithelial cells~\cite{Suh_2025}.
We used a 24-min stimulation train (1 s pulses every 90 s) with a time-averaged irradiance of $\sim$1.7 mW~cm$^{-2}$, an integrated dose $\sim$4.5--6.5$\times$ gentler than the 3~hr base regimen previously applied in \cite{Suh_2025}.
2P-OPM imaging ($\lambda_{\mathrm{Ex}}=1030$~nm; $\lambda_{\mathrm{Em~peak}}\approx603$~nm) reveals light-evoked subcellular reorganization: pre- and post-stimulation volumes show the appearance of new puncta and the brightening of pre-existing ones with high contrast throughout the tissue depth (Fig.~\ref{fig:main6}a,b).
Clustering was quantified by a spatial heterogeneity index $H$ (see Methods, Eq.~\eqref{eq:Hdef}) across four regions of interest (ROI).
Under blue stimulation, $H$ increases by $\sim\!45\%$ across all ROI pairs (Fig.~\ref{fig:main6}c); difference maps localize these gains to receptor-enriched membrane assemblies (Fig.~\ref{fig:main6}d).
These observations match the OptoEGFR mechanism, in which blue light triggers receptor oligomerization and activation \cite{Suh_2025}.
Together with the lower stimulation dose, the spatially localized gains indicate that 2P-OPM both delivers optogenetic control and quantifies subtle, molecular-scale processes in living tissue.

\section*{Discussion}
\label{sec:discuss}
2P-OPM brings the nonlinear advantages of multiphoton excitation to high-NA single-objective light-sheet microscopy, enhancing high-resolution imaging at depth with conventional sample mounting.
This reshapes the photon-budget constraints in multiphoton microscopy:
By exciting an entire plane at reduced illumination NA, the photon budget is spread across many voxels, extending total illumination time per voxel and delivering over an order-of-magnitude lower peak intensity at the same average power and axial resolution to 2P point scanning.
This reduction in peak intensity reduces susceptibility to higher-order nonlinear damage and allows higher tolerable average powers, enabling faster, gentler volumetric imaging than conventional 2P.

Our instrument delivers raw resolutions of \(\sim\!300\)~nm laterally and \(\sim\!650\)~nm axially, the highest reported for multiphoton OPM.
In multicellular contexts, we observe $\sim\!2\times$ higher contrast integrated over the full illumination depth, better transmission of spatial frequencies, and stronger optical sectioning than 1P-OPM at matched detection NA, yielding clearer cellular and subcellular images in scattering volumes.
In low-SNR conditions, we find $>\!2\times$ recovery in volumetric resolving power, with the largest gains along the axial direction, where scattering-induced blur most degrades 1P performance.

These gains and the extra bandwidth in the photon budget enable faster, more multiplexed, or more sensitive subcellular imaging \textit{in vivo}.
In living fly embryos, we can record 4D transcriptional dynamics across hundreds of individual nuclei at speeds exceeding conventional 2P point scanning by more than fivefold, simultaneously capture heterogeneous transcription-factor distributions and active gene loci, and resolve single mRNA molecules emerging from transcription sites and track their motion within the nucleus and syncytium.
Beyond imaging, the same NIR excitation also permits visible optogenetic perturbation without crosstalk.
OptoEGFR activation required $4.5\times$ gentler integrated irradiance and $3\times$ shorter stimulation than previously reported \cite{Suh_2025}, suggesting that prolonged perturbation-measurement durations at high spatiotemporal resolution are possible.

Because overall 3D image quality is limited by the poorest-resolution axis, PSF anisotropy matters.
As in other high-NA OPMs, lateral asymmetry from the remote-focus tilt is modest, while axial resolution is typically $\sim$2--5$\times$ worse \cite{Yang_2019,Sapoznik,Yang_2022}.
Thickening the light sheet extends propagation length and field of view but degrades $z$ resolution and sectioning; thinning it does the opposite.
A 0.5-\si{\micro\meter}-thin Bessel-like sheet approaches isotropy and triples the field of view (Supplementary Figs.~\ref{fig:layer-cake_sim} and~\ref{fig:2P-OPM_beams}).
The Bessel-like sheet spreads its power along a longer line of incoherent beamlets \cite{ChenChakraborty_2020}, so each point is dimmer and excites less in-focus 2P signal than a Gaussian sheet at the same average power; its side lobes also raise background within the field, which the 2P nonlinearity holds well below the 1P level but above that of a 2P Gaussian sheet (Supplementary Fig.~\ref{fig:2P-OPM_Bessel}).

The high-NA OPM train (specifically O1, tube lenses, and scan optics) also reduces NIR throughput and accumulates group-delay dispersion, broadening ultrafast pulses and diminishing 2P signal at fixed average power. 
Both Gaussian and propagation-invariant beams are affected, with the latter more penalized because side-lobe power lowers main-lobe intensity.
Consequently, our live single-molecule mRNA imaging was confined to a fixed plane and limited to $\sim$50 time points before molecules moved out of focus or were lost to photobleaching.

Our current 2P-OPM design leaves room for improvement.
Replacing GM-$x$ with a resonant galvo in time-averaged mode would illuminate each voxel multiple times within one exposure, spreading the dose in time (lower per-pass peak dose) while preserving the integrated dose per frame.
This has been reported to reduce photobleaching and phototoxicity at a given average power and, when shot-noise-limited, to maintain SNR at higher sampling rates \cite{Madaan_2021,Zhou_2022}.
A higher-transmission, visible--NIR-coated objective and dispersion compensation to restore bandwidth-limited pulses \cite{Chauhan_2010} should further boost 2P signal rates; mirror-based scan optics \cite{Gorlitz_2014,Sirinakis_2024} could further reduce group-delay dispersion and improve light throughput compared to refractive elements.

Even with 2P-OPM’s benefits in optically heterogeneous tissue, depth-dependent aberrations distort excitation and detection wavefronts.
Adaptive optics can recover near-diffraction-limited resolution and peak signal \cite{Liu_2018,Zhang_2023,McFadden_2024}, but adds hardware, calibration, and local, field-dependent corrections.
A single correction is at best partial; tiled corrections recover resolution only by sacrificing the very speed and large-field benefits that motivate light-sheet imaging in the first place \cite{Liu_2018}.

2P-OPM itself is also exacting and expensive to build and operate.
The remote-focus system, two microscopes back-to-back to form a remote intermediate image with equal lateral and axial magnification \cite{Botcherby_2007}, requires specialized optical expertise.
OPM then adds a third, tilted microscope to relay and magnify an oblique plane from the remote intermediate image to the camera \cite{Dunsby_2008}.
Room-temperature fluctuations of $\sim$1--2\si{\celsius} can induce $\sim$2--3~\si{\micro\meter} axial drift in the remote-focus space \cite{Nguyen_2025}, placing strict tolerances on O2--O3 and making drift mitigation essential to maintain long-term alignment.
Ultrafast lasers for 2P excitation add cost, and demand precise co-alignment of all optical paths.
Wavelength mixing can enable simultaneous 2P excitation of three fluorophores with just two synchronized beams; however, it requires tight spatiotemporal overlap and dispersion matching \cite{Mahou_2012,Mahou_2014}.
Reducing complexity and cost will be critical for reliability, scalability, and broader access \cite{Xu_2013}.

Fluorophore photophysics and photodamage thresholds ultimately bound  performance.
2P cross-sections are typically smaller than for 1P, lowering signal per unit average power; at matched voxel size and exposure, 1P and 2P achieve equivalent volumetric speeds, but the excitation modality determines the useful information content. For dim, thin, non-scattering specimens where background is negligible, maintaining adequate SNR at high voxel rates with 2P requires higher average power, which is ultimately bounded by photodamage thresholds, effectively limiting 2P volumetric speed relative to 1P in this regime.
We quantify this 1P--2P tradeoff in fixed tissue (Figs.~\ref{fig:main3} and \ref{fig:main4}), not live, as done for 2P scanned light-sheet microscopy~\cite{Truong_2011}.

One-color imaging permits selection of the most photostable fluorophore, whereas multicolor often forces incorporation of less photostable labels.
Continued refinement of bright, photostable 2P and NIR probes, including yellow- and red-shifted genetically encoded calcium indicators \cite{Mohr_2020,Yokoyama_2024}, rhodamine-based small molecules \cite{Grimm_2021}, and emitters in the NIR-II window \cite{Wang_2024}, will be key to faster, gentler, and highly multiplexed multiphoton imaging.
Complementary triplet-state strategies, such as NIR co-illumination (to drive reverse intersystem crossing) \cite{Ludvikova_2024}, can further reduce photodamage, extend measurement duration, and accelerate frame rates at the same SNR.

2P-OPM records high-speed, 4D molecular movies in intact multicellular systems.
Crucially, it requires no changes to standard sample mounting, lowering the barrier to wider adoption.
These capabilities usually trade against each other.
Combined, they let a single experiment map fast microscopic interactions and the macroscopic function they produce \cite{Betzig_2025,Bialek}.
Next-generation microscopes can incorporate structured illumination to unlock super-resolution in thick, densely fluorescent specimens \cite{Gao_2014,Gustafsson_2008}.
By relaxing resolution for field of view, whole-organoid imaging of subcellular dynamics also becomes feasible, enabling 4D single-cell tracking during self-organization \cite{McNamara_2024,Bennabi_2025}.
The same capabilities translate naturally to high-content drug screening on patient-derived organoids and 3D pathology for clinical applications \cite{Hsieh_2025,Liu_2023}.



\section*{Methods}

Our research complies with all relevant ethical regulations.
This study involved no vertebrate animals and no human participants, and required no approval by an institutional animal care and use committee or institutional review board.

\label{sec:2P-OPM_methods}

\subsection*{Theory and numerical simulations}
\label{sec:2P-OPM_sim_methods}

\subsubsection*{Detection point spread function}
We generated the 2P-OPM detection PSF by considering the primary (O1), secondary (O2), and tertiary (O3) objective lenses with the model of Yang \textit{et al}.~\cite{Yang_2022}, using code written in Python and accessible on \href{https://github.com/royerlab/daxi/tree/master/simulation_code}{GitHub}.
A plane wave passes through O2 and the pupil function is estimated after propagating out of O3.
The PSF is then simulated from this pupil function.
O2 (NA = 0.95) and O3 (NA 1.0) are air ($n = 1.0$) and solid immersion ($n = 1.515$) objectives, operating either in straight transmission (i.e., angle between O2 and O3 optical axes = $0^\circ$; Supplementary Fig.~\ref{fig:2P-OPM_sim}a) or with O3 tilted by a $45^\circ$ angle with respect to O2's optical axis (Supplementary Fig.~\ref{fig:2P-OPM_sim}b).
With a $45^\circ$ angle between O2 and O3, the effective pupil function of the remote-focusing system shows a compression of light in the tilted ($y$) direction (Supplementary Fig.~\ref{fig:2P-OPM_sim}b).

To obtain FWHM measures of the theoretical resolution, the PSFs are fit to a 1D Gaussian along the three principal axes (Supplementary Fig.~\ref{fig:2P-OPM_sim}).
The tilted PSF is first rotated to have its principal axes along the $x$-, $y$- and $z$-axes and then fit with a 1D Gaussian to get the FWHM.
In both the straight and tilted cases, the FWHM is further scaled by 1.515× to account for the magnification between O1 and O2, which is the ratio of the indices of refraction between O1 and O2, satisfying both the sine and Herschel condition to achieve perfect (i.e., aberration-free) remote refocusing~\cite{Botcherby_2007}.

\subsubsection*{Two-photon Bessel-like beams}
\label{sec:2P-cake_methods}

We generated 2P Bessel-like excitation beams by placing a pupil-plane layer-cake phase mask~\cite{Abrahamsson_2006,ChenChakraborty_2020,ExD_KKD} in a plane conjugate to the objective back pupil (Supplementary Fig.~\ref{fig:layer-cake_sim}).
The mask consists of concentric glass discs; adjacent zones introduce an optical path difference $\delta=t(n-1)$ (layer thickness $t$, refractive index $n$).
For ultrafast sources (pulse durations $\sim$100–200 fs; coherence length $\lesssim$ a few tens of \si{\micro\meter}), choosing $\delta \gg$ coherence length makes the annular sub-apertures mutually incoherent.
Each zone then forms its own elongated focus; their intensities add incoherently to yield an effective extended depth-of-focus excitation profile with near-Gaussian lateral width but extended axial extent.

We computed excitation foci using the vectorial Debye diffraction integral with a Fourier-domain implementation~\cite{Leutenegger_2006,ChenChakraborty_2020}, the same code as our previous extended depth-of-field light-sheet work~\cite{ExD_KKD}.
A continuum of wavevectors transmitted by an objective lie on a spherical cap with a magnitude $k=n/\lambda$.
Lateral and axial feature sizes map to the wavevectors along $k_x$ and $k_z$, respectively.
The set of $k_x$ wavevectors is given by $2k\sin(\alpha_2)$; the set of $k_z$ wavevectors is given by $k(\cos(\alpha_1) - \cos(\alpha_2))$.
For a conventional (central) sub-aperture using the full spherical cap ($\alpha_1{=}0$), the axial extent is
\begin{equation}
\label{eq:dz}
d_z=\frac{\lambda}{n\!\left(1-\cos(\alpha_{2})\right)}.
\end{equation}
Annular sub-apertures ($\alpha_1{\neq}0$) produce Bessel-like foci with axial extent
\begin{equation}
\label{eq:dz_Bessel}
d_z=\frac{\lambda}{n\!\left(\cos(\alpha_{1})-\cos(\alpha_{2})\right)}.
\end{equation}
Using $\alpha=\arcsin(\mathrm{NA}/n)$ gives
\begin{equation}
\label{eq:dz_Bessel_NA}
d_z=\frac{\lambda}{n\!\left[\cos\!\left(\arcsin(\frac{\mathrm{NA}_1}{n})\right)-\cos\!\left(\arcsin(\frac{\mathrm{NA}_2}{n})\right)\right]}.
\end{equation}

Because the back pupil diameter is $2f\,\mathrm{NA}$ (where $f$ is the objective back focal length), Eqs.~\eqref{eq:dz}–\eqref{eq:dz_Bessel_NA} convert NAs directly to the corresponding mask zone radii when properly accounting for the relay magnification between them.
We determine zone sizes iteratively so each zone yields the same $d_z$: choose the central (low-NA) zone to set the target $d_z$ via Eq.~\eqref{eq:dz}, then pick the next annulus $(\mathrm{NA}_1,\mathrm{NA}_2)$ from Eq.~\eqref{eq:dz_Bessel_NA} so its Bessel-like focus has the same $d_z$, and repeat outward until the desired number of zones (or maximum NA) is reached.
This construction yields the expected near-linear scaling of the depth of focus with the number of annuli.

For each zone we compute the complex electric field $E(\mathbf{r})$, form the 2P excitation foci as $|E|^{4}$, and then sum incoherently across zones to obtain the axially elongated 2P focus.
The central zone produces a Gaussian-like focus (Eq.~\ref{eq:dz}); outer annuli produce Bessel-like foci (Eq.~\ref{eq:dz_Bessel_NA}).
The summed extended focus maintains lateral width close to that of the highest NA zone while extending the axial extent set by the design $d_z$ (see Supplementary Figs.~\ref{fig:layer-cake_sim} and~\ref{fig:layer-cake_design}).
Wavelengths and NAs closely matched the experimental parameters: $\lambda_{\mathrm{exc}}=900$ \si{\nano\meter}, $\text{NA}_{\mathrm{exc}}=0.5$, $n=1.515$, and $L=4$, where $L$ is the number of annuli; this yielded a depth of focus of 10 \si{\micro\meter}; the conventional high-NA Gaussian beam was identical, except with a depth of focus of 2.5 \si{\micro\meter}.
We experimentally validated these theoretical predictions by imaging a solution of rhodamine (Supplementary Fig.~\ref{fig:2P-OPM_beams}), and characterized the performance by imaging a DAPI-stained fruit fly embryo (Supplementary Fig.~\ref{fig:2P-OPM_Bessel}).

\subsection*{Microscope optics}
\label{sec:2P-OPM_optics}
Refer to Fig.~\ref{fig:main1}a for the beam paths and key components.
Table~\ref{tab:parts} contains a list of acronyms and key parts used to construct the microscope;
a complete parts list is in the computer-aided design (CAD) model.
Supplementary Fig.~\ref{fig:2P-OPM_CAD+photo} shows a CAD and photos of the 2P-OPM instrument.

Two Ti:Sapphire lasers and an ytterbium-based laser are used for 2P excitation (red path).
Intensity modulation for each ultrafast laser source is achieved using Pockels cells (PC1-PC3). The NIR beams are combined using broadband and dichroic mirrors (DM7 and DM8): the first Ti:Sapph (750--850~nm) reflects off DM7, the second (850--980~nm) transmits through DM7 and reflects off DM8, and the ytterbium laser (1030~nm) transmits through both, spanning 750--1030~nm. The combined beam is expanded to a 12-mm diameter, with an adjustable iris (I) downstream to tune the illumination NA. A layer-cake phase mask (MSK) can optionally generate a Bessel-like focus (see Methods above).
CW laser sources at 405 nm (purple), 488 nm (blue), and 561 nm (yellow) are used for linear 1P excitation.

These beams are collimated and expanded by lenses L5$\textsubscript{1-3}$ and L6$\textsubscript{1-3}$ to a 1.3-mm diameter and combined via broadband and dichroic mirrors (DM5 and DM6) into a single beam. This beam passes through a half-wave plate ($\lambda$/2$\textsubscript{VIS}$) and acousto-optic tunable filter (AOTF) for modulation (blue path), then expands 7.5× via lenses L3 and L4. The illumination NA is controlled with an iris (I) before the beam is merged with the NIR light path using DM4.

Lens L2 focuses the combined visible and NIR co-linear beam onto resonant galvo (RG).
The beam is then conjugated to scanning galvo mirrors GM-$x$ and GM-$y$, and the rear focal plane of the primary objective O1 in sequence through lens L1, tube lens T2 and scan lens SL2, and SL1 and T1.
For proper conjugation and positioning of the galvos, RG and GM-$x$ are mounted on 2D translation stages; GM-$y$ is mounted on a 3D translation stage.
L2 through GM-$x$ sit on a breadboard bolted to a $y$-translation stage (Supplementary Fig.~\ref{fig:2P-OPM_CAD+photo}d).
DM1 directs the visible and NIR light to O1; translation of the $y$ stage offsets the beam from center so that it intersects the specimen obliquely.
3D motion-control stages translate the sample and an environmental chamber provides closed-loop control of temperature, humidity and gas concentration.

Fluorescence emission is collected through O1 and propagated through the remote-focus optics (O2 and O3) to form an image of the oblique plane within the specimen.
DM2, DM3, and a broadband mirror split the fluorescence into three spectral emission channels, which are focused onto Sci-Cam1, Sci-Cam2, and Sci-Cam3 with T3$\textsubscript{1-3}$ (Supplementary Fig.~\ref{fig:spectra}).
O1--03 and Sci-Cam1--3 are all fixed on 2D translation stages for precise alignment.
O3 is further mounted to a piezoelectric collar and kinematic mount (with tip/tilt) for precise positioning and alignment to the 1P and 2P light sheet (Supplementary Fig.~\ref{fig:2P-OPM_CAD+photo}e).

Flip mirror FM directs the multicolor excitation beam to a transmitted wide-field illumination mode, which includes a movable dichroic mirror (MDM) and lenses L7-L8 mounted diascopically above O1.
This mode also serves as a practical alignment laser for the light microscope; specifically, the correct distance between each pair of lenses in the system is confirmed with a shearing interferometer.
Removing MDM permits a beam splitter (BS) to inject 455 nm and/or 625 nm LED light for bright-field illumination or photoactivation.
Movable mirror MM1, positioned between SL2 and T2, diverts detected light to a wide-field inspection camera (WF-Cam) for low-magnification view-finding and macroscopic imaging (see Supplementary Fig.~\ref{fig:WF-Cam_photo+embryo} and Supplementary Movie~\href{https://kdizon.github.io/2P-OPM/movies/WF-embryo_movie.mp4}{1}), using relay lens R1 ($f = 100$ mm) and T5.


\squeezetable
\begin{table}[htbp]
\caption{\textbf{Key optical components and acronyms used in Fig.~\ref{fig:main1}.}}
\label{tab:parts}
\footnotesize
\setlength{\tabcolsep}{4pt}\renewcommand{\arraystretch}{1.05}
\begin{tabularx}{\columnwidth}{@{}l>{\raggedright\arraybackslash}X@{}}
\toprule
\textbf{Acronym} & \textbf{Part/description (vendor, model)} \\
\midrule
O1 & Primary objective (Olympus $60\times$, NA 1.49, oil immersion) \\
O2 & Secondary objective (Olympus $40\times$, NA 0.95, air) \\
O3 & Tertiary objective (Applied Scientific Instrumentation $40\times$, NA 1.0, glass) \\
T1--T2 & Tube lens (Thorlabs AC508-180-AB-ML)\\
T3 & Tube lens (Edmund Optics 49-287)\\
T5 & Tube lens (Thorlabs AC254-050-A-ML) \\
SL1--SL2 & Scan lens (Plössl; composed of two $f = 200$ mm VIS-NIR achromatic doublets; Edmund Optics 49-392) \\
GM-$y$ & $y$ galvo for OPM galvo scanning/descanning (Thorlabs GVS011)\\
MM1 & Movable mirror 1 \\
RL1 & Relay lens 1 ($f=100$\,mm) \\
WF-Cam & View-finding camera for wide-field imaging through O1 (Basler ace acA2040-55um)\\
DM1 & Dichroic mirror 1 (Semrock Di01-R405/488/561/635/800-t1-25$\times$36) \\
DM2 & Dichroic mirror 2 (Semrock Di03-R488-t3-50.8$\times$55) \\
DM3 & Dichroic mirror 3 (Semrock Di03-R561-t3-50.8$\times$55) \\
Sci-Cam1--3 & PCO edge4.2 Camera Link HS\\
L1--10 & Lenses with labeled focal lengths (includes diascopic L7--L8) \\
GM-$x$ & $x$ galvo for scanning beam into sheet (Cambridge Technology) \\
RG & Resonant galvo (sheet wobble in $xy'$ to reduce shadowing; Cambridge Technology CRS04) \\
DM4 & Dichroic mirror 4 (Semrock FF665-Di02-25$\times$36) \\
FM & Flip mirror \\
I & Iris, adjustable (illumination NA control) \\
AOTF & Acousto-optic tunable filter (continuous-wave laser modulation; AA Opto-Electronic AOTFnC-400.650-CPCh-TN; 400-650 nm, 2.5 mm aperture)\\
DM5 & Dichroic mirror 5 (Semrock FF665-Di02-25$\times$36) \\
DM6 & Dichroic mirror 6 (Semrock FF665-Di02-25$\times$36) \\
MDM & Movable dichroic mirror (diascopic module above O1) \\
BS & Beamsplitter (LED injection for bright-field) \\
DM7 & Dichroic mirror 7 (Semrock FF850-Di01-25$\times$36) \\
DM8 & Dichroic mirror 8 (Semrock FF980-Di01-25$\times$36) \\
PC1--PC3 & Pockels cells (ultrafast laser modulation; ConOptics M350-80-02 \&M302RM) \\
$\lambda/2$ & Half-wave plate, where the subscripts VIS and NIR refer to the visible and near-infrared wavelengths \\
MSK & Layer-cake phase mask (Bessel-like excitation; LightMachinery, custom; see Supplementary Fig.~\ref{fig:layer-cake_design}) \\
Ti:Sapph & Ti:Sapphire lasers (Coherent Chameleon Ultra I; 680–1040 nm, 2.9 W @800 nm, 80 MHz rep rate, 140 fs pulse duration) \\
Yb & Ytterbium-based laser (Light Conversion, FLINT-7W-SH; 1030 nm, 7 W, 11.4 MHz rep rate, sub-120 fs pulse duration) \\
405 nm & 405 nm laser (Coherent OBIS LX 1185050, 200 mW) \\
488 nm & 488 nm laser (Coherent Sapphire LP, 300 mW) \\
561 nm & 561 nm laser (Coherent Sapphire LP, 300 mW) \\
455 nm & 455 nm LED for bright-field and/or optogenetic stimulation (Thorlabs M455L4; 1445 mW) \\
625 nm & 625 nm LED for bright-field and/or optogenetic stimulation (Thorlabs M625L4; 920 mW) \\
\bottomrule
\end{tabularx}
\end{table}

\subsection*{Microscope control}
\label{sec:2P-OPM_control}
All acquisitions ran on a Colfax ProEdge WX3400–class workstation equipped with an Intel Xeon W7-3445 processor operating at 2.6 GHz with 20 cores and 40 threads, 192 GB DDR5-4800 ECC RAM, and a Samsung PM9A3 3.84 TB M.2 NVMe SSD.
Data display and file writing were handled in \textmu-Manager~2.0~\cite{Edelstein_2014}.
A custom Python control module (fastMC) programmed a National Instruments X\mbox{-}Series DAQ (NI\mbox{-}PCIe\mbox{-}6321 with NI\mbox{-}BNC\mbox{-}2110 breakout) via the \texttt{nidaqmx} API to generate synchronized hardware triggers (TTL pulses) and analog waveforms that control four cameras (Sci-Cam1--3, plus WF-Cam), the $y$-scan galvo (GM-$y$), the lasers (AOTF/Pockels), and the bright-field/optogenetic LEDs.
All DAQ tasks (counters, digital lines, analog outputs) were precomputed and hardware-clocked for the acquisition duration within device limits.
fastMC incorporates and adapts elements from the acquisition scripts of~\cite{Yang_2022}.

Sci-Cam1--3 were run in rolling-shutter mode with external exposure control, in which each TTL pulse defines both the start and the duration of an exposure (pre-defined settings are ignored).
fastMC uses a primary counter to mark frames (2D) or $y$-stacks (3D), and start GM-$y$ (if 3D imaging) and the LED (both driven by pre-determined signal trains).
A gated secondary counter emits the exposure-controlling pulse train accepted by Sci-Cam1--3.
The exposure pulse has duty cycle $\sim$0.98 and a period equal to the requested exposure, which ensures trigger acceptance across ROIs/readout heights and enables maximal full-sensor rates (e.g., 100 fps).
See Supplementary Fig.~\ref{fig:2P-OPM_timing} for the timing diagram and the schematic of device-DAQ connections.

GM-$y$ was driven from a $\pm 5$\,V waveform (stepped once per frame during volumetric imaging) so the sample remained effectively stationary during each rolling-shutter exposure.
Laser power was controlled by an analog output to the AOTF/Pockels with a concurrent TTL blanking line automatically emitted by any one of Sci-Cam1--3; excitation was on only during exposure windows and off during readout, inter-frame gaps, and galvo retrace.
LED illumination offers two modes: a stack-synchronized fraction mode (LED on for a chosen fraction of each $y$-stack) and an independent timed mode (a user-specified on window starting with acquisition).

fastMC was used specifically for high-speed 2D/3D acquisitions requiring deterministic hardware timing and sustained multi-camera recording without drops; routine preview and slower measurements were run directly in \textmu-Manager.
Multi-dimensional data were saved by \textmu-Manager (NDTiff/TIFF series) directly to a network-attached storage (NAS) server (Synology DS1821+ 8-Bay Diskstation) connected via 10 GBe Ethernet to the acquisition PC.

\subsection*{Sample preparation}
\label{sec:2P-OPM_sample}

\subsubsection*{Cleaning and sterilization of coverslips}
Glass coverslips (\#0; Thorlabs, CG00K) were cleaned following an alkaline–peroxide protocol~\cite{Gao_2014}. A fresh 150\,mL solution of hydrogen peroxide (H$_2$O$_2$, 50\%), ammonium hydroxide (NH$_4$OH, 30\%), and H$_2$O was prepared at a 1:1:5 volume ratio.
Coverslips were immersed and soaked for 12 hr at room temperature, then rinsed thoroughly in deionized water (5$\times$) and air-dried in a dust-free container.
Immediately before use, coverslips were sterilized by a brief methanol rinse followed by flaming and allowed to cool completely.

\subsubsection*{Fluorescent beads}
Clean 5–8 mm$^{2}$ rectangular coverslips were coated with 0.1\,mg\,mL$^{-1}$ poly-D-lysine (30 \textmu L per coverslip), air-dried for 15 min, and rinsed in water.
Sub-diffractive fluorescent beads (50 nm diameter; Dragon Green, Bangs Laboratories, FSDG001) were diluted $\sim\!10^{6}\times$ in ethanol; 30 \textmu L was deposited on the coated coverslip and allowed to evaporate and dry.
A small drop of immersion oil (Olympus) was applied to immobilize and index-match the dried beads for PSF measurements.
Samples were stored in the dark at room temperature and reused until bleached.

\subsubsection*{mESC culture}
Mouse embryonic stem cells (mESCs) 129/SvEv (EmbryoMax, MilliporeSigma, CMTI-1), obtained from the Gregor laboratory, were cultured in 0.1\% gelatin-coated six-well plates in a humidified incubator (5\% CO$_2$, \SI{37}{\celsius}).
Cells were maintained in basal growth medium comprising Glasgow Minimum Essential Medium (G-MEM, 11710-035) supplemented with 10\% ESC qualified fetal bovine serum (R\&D Systems, S10250), 1:100 GlutaMAX (Gibco, 35050–061), 1:100 MEM non-essential amino acids (Gibco, 11140-050), 1 mM sodium pyruvate (Gibco, 11360–070), 100 $\mu$M 2-mercaptoethanol (Gibco, 21985-023), and 1\% penicillin-streptomycin (10,000 units/mL, Gibco, 15140-122).
The basal growth medium was additionally supplemented with 1000 units/mL LIF (Millipore Sigma, ESG1107), 2 $\mu$M PD 0325901 (Tocris, 4192), and 3 $\mu$M CHIR 99021 (Tocris, 4423) to maintain pluripotency.
The medium was replaced every day and cells were passaged every other day using Trypsin (Gibco, 12605-010) dissociation and a seeding density of $3\times10^5$ cells per well.
The mESC line was not authenticated.

\subsubsection*{Gastruloid culture}
mESC-derived gastruloids were generated as described in~\cite{Beccari_2018}. In-house N2B27 medium was freshly prepared maximally one week in advance of the experiments as follows: 1:1 mix of DMEM/F12 +GlutaMax (Gibco, 10565-018) and Neurobasal (Gibco, 21103-049), supplemented with 100 $\mu$M 2-mercaptoethanol, 1:100 N2 (Gibco, 17502-048), 1:50 B27 (Gibco, 17504-044) and 1\% penicillin-streptomycin (10,000 units/mL).
Cell seeding for gastruloid generation was performed manually using a multipipette and the cell counts were determined with an automatic cell counter (Countess 3 Automated Cell Counter, Invitrogen).
For cell dissociation from adherent mESC cultures, cells were treated with Trypsin and immediately resuspended in N2B27 medium.
The dissociated cells were seeded into Costar Low Binding 96-well plates (Costar, Corning, 7007) with 300 cells per 40 $\mu$L per well.
Gastruloids were left untouched for 48 hr to facilitate aggregation.
At 48 hr, the resulting spheroids were subjected to a 24-hr pulse of Wnt agonist through the addition of 150 $\mu$L of 3 $\mu$M CHIR 99021 (Chiron) in N2B27 to each well.
Subsequently, 150 $\mu$L of the medium was replaced every 24 hr until gastruloid collection at 96 or 120 hr.

\subsubsection*{Immunofluorescent staining of gastruloids}
Gastruloids were collected from the well plate, pooled in a 15 mL falcon tube, and washed once with PBS with Mg${2+}$ and Ca${2+}$ (PBS++, Gibco, 14040-133).
After gastruloids sank to the bottom of the tube, PBS++ was removed and gastruloids were subsequently fixed in 10~mL 4\% paraformaldehyde solution (PFA, Thermo Scientific Chemicals, 30525-89-4) for 2 hr at \SI{4}{\celsius}.
Afterwards, gastruloids were washed twice with 10 mL PBSF (10\% FBS in PBS++) for 15 min at room temperature, resuspended in 1 mL in PBS++ and stored at \SI{4}{\celsius} (for several weeks).
For the immunofluorescent staining, gastruloids were permeabilized in 10 mL PBSFT (10\% FBS and 0.03\% Triton  in PBS++) and incubated for 2 hr at room temperature.
Gastruloids were then incubated in 0.5 mL PBSFT containing 4',6-diamidino-2-phenylindole (DAPI) and primary antibody overnight at \SI{4}{\celsius} (see Table \ref{tab:antibodies} for details on antibodies and concentrations).
The next day, gastruloids were washed three times with 10 mL PBSFT for 30 min each at room temperature and subsequently incubated in 0.5 mL PBSFT containing secondary antibody and DAPI overnight at \SI{4}{\celsius} (Table \ref{tab:antibodies}).
Gastruloids were subsequently washed twice in 10 ml PBSFT and once in PBS++ at room temperature for 30 min each.
All incubations and washes were performed under nutation.
For gastruloid mounting, all excess PBS++ was removed from the tube and replaced with 200 $\mu$L mounting medium composed of 50:50 Aqua-Poly Mount (Polysciences, 18606-20) and PBS++.
Gastruloids in mounting medium were then transferred to a round 35-mm glass-bottom dish (No. 0 coverglass, MatTek Corporation, P35G-0-14-C), covered with a coverglass, and sealed with nail polish.

\squeezetable

\begingroup
  \renewcommand\tabularxcolumn[1]{m{#1}}
  \setlength{\tabcolsep}{3pt}
  \renewcommand{\arraystretch}{1.08}
\begin{table}[htbp]
  \centering
  \caption{\textbf{Primary and secondary antibodies and fluorescent dyes.}}
  \label{tab:antibodies}
  \footnotesize

  \begin{tabularx}{\columnwidth}{@{}%
    >{\raggedright\arraybackslash}m{0.2\columnwidth}   
    >{\centering\arraybackslash}m{0.12\columnwidth}     
    >{\centering\arraybackslash}m{0.17\columnwidth}     
    >{\centering\arraybackslash}m{0.25\columnwidth}     
    >{\centering\arraybackslash}X     
  @{}}
  \toprule
  \textbf{Antibody} & \textbf{Species} & \textbf{Catalog} & \textbf{Vendor} & \textbf{Dilution} \\
  \midrule
  BRA                & Goat   & AF2085     & R\&D Systems  & 1:100 \\
  FOXC1              & Rabbit & EPR20678   & Abcam         & 1:500 \\
  SOX2               & Rat    & 14-9811-82 & eBioscience   & 1:200 \\
   \makecell[l]{anti-goat\\AF-488}   & Donkey & A-11055    & Invitrogen    & 1:500 \\
   \makecell[l]{anti-rabbit\\AF-546} & Donkey & A-11010    & Invitrogen    & 1:500 \\
  \makecell[l]{anti-rat\\AF-647} & Donkey & A78947 & Invitrogen & 1:500 \\
  DAPI               & —      & D3571      & Invitrogen    & 1:500 \\
  \bottomrule
  \end{tabularx}
  \end{table}
\endgroup

\subsubsection*{smFISH probe synthesis}
Probes were designed computationally with a 25-nucleotide hybridization region with a melting temperature of at least 45\si{\celsius}, and specific 20-nucleotide forward primer and 30-nucleotide reverse primer with an incorporated T7 promoter.
Probe synthesis followed a similar protocol outlined in~\cite{Mateo_2021}.
Briefly, probes were ordered using an oligo library from IDT at 10\si{pmol}/oligo concentration. Probes were amplified with PCR and then treated with a \href{https://www.neb.com/en-us/products/e2050-hiscribe-t7-quick-high-yield-rna-synthesis-kit}{HiScribe T7 (E2050)} in-vitro-transcription (IVT) utilizing the incorporated T7 promoter.
A specific ATTO-conjugated primer (also from IDT) was added and incorporated into the probes following a reverse transcription reaction with the \href{https://www.thermofisher.com/order/catalog/product/K1652}{Maxima H Minus RT (K1652)} kit.
Resulting ssDNA probes were treated with RNAse A, clean, and stored at 4\si{\celsius}.

\subsubsection*{smFISH on fly embryos}
Oregon--R (Ore-R) fly stocks were maintained on standard cornmeal medium.
Embryos were fixed in 4\% formaldehyde and stored in methanol at -20\si{\celsius}.
smFISH targeted the exonic region of the \textit{even\mbox{-}skipped} gene (\textit{eve}) using ATTO 565- and ATTO 633-conjugated probes.
Hybridization was performed with 750 \si{ng} of conjugated probe in 5$\times$\,SSC buffer (Thermo Fisher Scientific, 15557044), 50\% formaldehyde, and 0.1\% Tween-20 at 37\si{\celsius} overnight.
Embryos were washed, mounted in ProLong Gold (Invitrogen), and allowed to cure overnight prior to imaging.

\subsubsection*{Genetic crosses for live-fly imaging}
To obtain embryos co-expressing Bcd and nascent \textit{hb} (Fig.~\ref{fig:main5}a), female virgins expressing Bcd–eGFP~\cite{Gregor_2007} were crossed to males carrying a maternal MS2 coat protein (MCP) line (\textit{nanos}$>$NLS--MCP--mCherry).
Female offspring (\textit{bcd--eGFP/+; nanos}$>$\textit{NLS--MCP--mCherry/+; +/+}) were then crossed to males homozygous for a \textit{hb}--MS2 reporter in which the \textit{hb} promoter drives a transcript containing 48 MS2 stem–loop repeats followed by the \textit{lacZ} coding sequence.
The 48$\times$ cassette was generated by duplicating a 24$\times$ MS2 cassette from \texttt{piB-hbP2PP2E-24MS2-lacZ-tub(3'UTR)} via NcoI and AvrII excision and insertion into the same vector digested with NcoI and SpeI (SpeI site upstream of the original cassette), resulting in tandem 48$\times$ MS2. The line was integrated by BestGene, Inc. into BDSC \#27388, genotyped, and transcription preliminarily confirmed by confocal microscopy.
Embryos from this cross were collected and used for live imaging.

For Fig.~\ref{fig:main5}b--e, female virgins of a line carrying nuclear markers and MCP (\textit{yw; His2Av--mRFP; nanos}$>$\textit{NLS--MCP--mNeonGreen})~\cite{Garcia_2013,Singh_2022} were crossed to wild-type (Oregon–R) males to reduce background fluorescence.
Female offspring (\textit{yw/+; His2Av--mRFP/+; nanos}$>$\textit{NLS--MCP--mNeonGreen/+}) were then crossed to males expressing the same \textit{hb}--MS2 reporter as above (48$\times$ MS2--\textit{lacZ}).
Embryos from this cross were collected and used for live imaging.

\subsubsection*{Fly mounting for live imaging}
Fly embryo preparation for live imaging was adapted from previous work~\cite{Garcia_2013,Chen_2025}.
A $\sim$ 2 × 2 \si{\centi\meter}$\textsuperscript{2}$ air-permeable membrane was coated with heptane glue, made by dissolving double-sided Scotch tape in heptane.
The membrane was left to dry for $\sim$10 min while embryos were collected, leaving an adhesive surface.
Flies were caged on agar plates for 2 to 2.5 hr, and the laid embryos were transferred to a piece of double-sided tape with a dissection needle, hand-dechorionated, and placed on the glued membrane, with the dorsal side up (i.e., facing the objective lens).
After mounting, we immersed embryos in Halocarbon 27 oil (Sigma) and sandwiched the specimen between the membrane and a clean coverslip (see \textit{Cleaning and sterilization of coverslips} section above).
Excess oil is removed via lens tissue wipes (Thorlabs, MC-5) as needed.

\subsubsection*{OptoEGFR cell culture}
RPE-1 cells expressing the OptoEGFR system were used from Suh, Thornton \textit{et al}.~\cite{Suh_2025}; the parental line is hTERT RPE-1 (female).
OptoEGFR was introduced by lentiviral transduction of a construct encoding an N-terminal membrane localization tag fused to the OptoDroplet light-inducible clustering system (FUS$^{\mathrm{N}}$–FusionRed–Cry2) and the cytosolic domains of EGFR, as described in~\cite{Suh_2025}.
Cells were cultured in DMEM/F12 (Gibco, 11320-033) supplemented with 10\% fetal bovine serum (R\&D Systems, 26140079), 1\% L-glutamine (Gibco, 25030-081), and 1\% penicillin-streptomycin (10,000~units/mL; Gibco, 15140-122).
All cells were maintained at \SI{37}{\celsius} and 5\% CO$_2$ in a humidified incubator.
RPE-1 cells were tested to confirm the absence of mycoplasma contamination.
The RPE-1 line was authenticated by STR profiling.

For imaging, glass-bottom plates (No.~1.5 coverglass, CellvIs, P96-1.5H-N) were coated with \SI{10}{\micro\gram\per\milli\liter} fibronectin dissolved in PBS at \SI{37}{\celsius} for a minimum of \SI{30}{\minute}.
Cells were then seeded on glass-bottom 96-well plates at $\sim$4$\times$10$^4$ cells per well 24 hr before imaging.
To enhance adhesion, \SI{100}{\micro\liter} of cell suspension was first added per well and briefly centrifuged (30 s); after adhesion, an additional \SI{100}{\micro\liter} of full medium was added.
3 hr before imaging, the growth medium was replaced with serum-free starvation/imaging medium (DMEM/F12 with 1\% L-glutamine and 1\% penicillin–streptomycin; no FBS).

\subsection*{Imaging conditions}
Detailed imaging conditions for each experiment can be found in Supplementary Table~\ref{tab:imaging}.

\subsection*{Image processing and analysis}

\subsubsection*{Image transformation and reconstruction}
Both 2P- and 1P-OPM acquire volumes in an oblique (non-Cartesian) coordinate system (Fig.~\ref{fig:main1}c,d), so we transform stacks to conventional microscope coordinates for most visualization and analysis (Supplementary Fig.~\ref{fig:recon}a–c).
The scan axis is tilted (by~$45^\circ$) relative to the image plane; successive frames are therefore displaced laterally and axially, yielding a sheared stack.
We correct this by an affine transformation that deskews, rescales, and rotates the volume into conventional microscope coordinates (Supplementary Fig.~\ref{fig:recon}c).
We calculate a new coordinate vector for every single voxel within a volume, so reconstruction can be computationally expensive, especially with time-series and/or multicolor data.
To accelerate our reconstruction pipeline, we load our data into a Python script with GPU acceleration (CuPy), which speeds processing by an order of magnitude over CPU-only code.

In conventional $z$-stacking, a 3D volume is assembled from successive 2D frames separated by an experimentally known $z$ step; specifying the inter-frame spacing relative to the in-plane (sensor) pixel size $a$ suffices to render voxels at the correct scale (Supplementary Fig.~\ref{fig:recon}a).
In OPM, by contrast, the scan direction is effectively oblique, so each frame advances in $z'$ and is laterally offset in $y'$ (Supplementary Fig.~\ref{fig:recon}b);
here $x$ and $y'$ denote the in-plane (camera) axes and $z'$ the axial direction along the detection tilt.
We account for both components of the scan step during deskewing/rotation (Eqs.~\ref{eq:skew-dist}–\ref{eq:scale-factor}) to obtain a reconstructed volume in conventional coordinates (Supplementary Fig.~\ref{fig:recon}c).

Let $y_{\mathrm{step}}$ be the physical scan increment per frame and $\theta$ the light-sheet angle (Fig.~\ref{fig:main1}c and Supplementary Fig.~\ref{fig:recon}b). The per-frame displacements along the oblique axes are
\begin{align}
d_{\mathrm{skew}} &= y_{\mathrm{step}}\cos\theta, \label{eq:skew-dist}\\
d_{\mathrm{scale}} &= y_{\mathrm{step}}\sin\theta. \label{eq:scale-dist}
\end{align}
Expressed in pixel units using the in-plane pixel size $a$ (\si{\micro\meter}/pixel),
\begin{align}
s_k &= \frac{d_{\mathrm{skew}}}{a}, \label{eq:skew-factor}\\
s_c &= \frac{d_{\mathrm{scale}}}{a}. \label{eq:scale-factor}
\end{align}

We apply three affine operations to homogeneous coordinates $\bigl[x\;\;y\;\;z\;\;1\bigr]^\top$, Skew ($K$), Scale ($S$), then Rotation ($R$), in that (non-commutative) order (Supplementary Fig.~\ref{fig:recon}c):
\begin{equation}
\begin{bmatrix} x \\[2pt] y \\[2pt] z \\[2pt] 1 \end{bmatrix}
=
R(\theta)\,
\underbrace{\begin{bmatrix}
1 & 0 & 0 & 0\\
0 & 1 & s_k & 0\\
0 & 0 & s_c & 0\\
0 & 0 & 0 & 1
\end{bmatrix}}_{\displaystyle S\!\cdot\!K}
\begin{bmatrix} x' \\[2pt] y' \\[2pt] z' \\[2pt] 1 \end{bmatrix}.
\label{eq:composite}
\end{equation}
The final rotation maps the oblique stack back to the conventional frame:
\begin{equation}
R(\theta)=
\begin{bmatrix}
1 & 0 & 0 & 0\\
0 & \cos\theta & -\sin\theta & 0\\
0 & \sin\theta & \cos\theta & 0\\
0 & 0 & 0 & 1
\end{bmatrix}.
\label{eq:rotation}
\end{equation}
(Pre- and post-translation matrices recenter the volume so rotations act about its origin; omitted here for clarity.)
After applying the affine transform, we resample by trilinear interpolation onto a regular Cartesian grid, yielding isotropic voxels ($a\times a\times a$); this deskewed and rotated volume is the reconstructed output (Supplementary Fig.~\ref{fig:recon}c, right).

\subsubsection*{PSF benchmarking}
Beads were detected and fit in PSFj to extract FWHM$_{\min}$, FWHM$_{\max}$, FWHM$_{z}$, and centroids \cite{Theer_2014}.
Lateral principal axes were mapped to microscope axes by assigning the shorter in-plane width to $x$ and the longer to $y$ (broadened by the O3 tilt).
Analysis covered the full field of view; we retained unsaturated, isolated beads meeting PSFj’s fit-quality criterion ($R^2 \geq 0.9$) and report mean $\pm$ SD and $N$.

\subsubsection*{Image contrast}
We quantified image contrast in Fig.~\ref{fig:main3}f following prior work~\cite{KKD_2020,Truong_2020,Madaan_2021}.
For an image with $N$ pixels and intensities $I_j$, the mean and standard deviation are
\begin{equation}
\label{eq:mean}
\bar{I}=\frac{1}{N}\sum_{j=1}^{N} I_j,\qquad
\sigma_I=\sqrt{\frac{1}{N-1}\sum_{j=1}^{N}\!\left(I_j-\bar{I}\right)^2}\, .
\end{equation}
We define contrast as the standard deviation divided by the mean (the coefficient of variation),
\begin{equation}
\label{eq:contrast}
\mathrm{Contrast}=\frac{\sigma_I}{\bar{I}}\, .
\end{equation}

We computed a depth-dependent contrast profile $\mathrm{Contrast}(y')$ from individual $xz'$ orthoslices as a function of $y'$ (distance from the glass interface) across the full illumination depth, and normalized it to the surface value, so that $\mathrm{Contrast}(0)=1$ for each modality.

The depth-averaged contrast is
\begin{equation}
\label{eq:avg_contrast}
\overline{\mathrm{Contrast}}=\frac{1}{L}\int_{0}^{L}\mathrm{Contrast}(y')\,dy'\, ,
\end{equation}
where $L$ denotes the analyzed depth span.
We evaluate $\overline{\mathrm{Contrast}}$ as the discrete mean of the sampled $\mathrm{Contrast}(y')$ values over the profile.

The depth-integrated contrast is
\begin{equation}
\label{eq:int_contrast}
\mathrm{Contrast}_{\mathrm{int}}=\int_{0}^{L}\max\!\big\{\mathrm{Contrast}(y')-1,\,0\big\}\,dy'\, ,
\end{equation}
computed from the sampled profile using the standard trapezoidal sum.
Reported improvements are ratios of $\mathrm{Contrast}_{\mathrm{int}}$ (or $\overline{\mathrm{Contrast}}$) for 2P-OPM relative to 1P-OPM under matched detection optics.

\subsubsection*{smFISH signal-to-noise ratio}
We quantified the SNR of the smFISH puncta shown in
Fig.~\ref{fig:main4}, adapting our prior approach \cite{ExD_KKD}. For each punctum,
\begin{equation}
\mathrm{SNR} = \frac{I_{\mathrm{spot}} - I_{\mathrm{bg}}}{\sigma_{\mathrm{bg}}\sqrt{N}},
\end{equation}

where $I_{\mathrm{spot}}$ is the integrated signal within an ovoid centered
on the punctum, $I_{\mathrm{bg}}$ is the integrated background from an
equivalent ovoid in a nearby region devoid of puncta,
$\sigma_{\mathrm{bg}}$ is the per-pixel standard deviation of that background,
and $N$ is the number of pixels in the ovoid, so that the background noise is
propagated over the same support as the integrated signal. We used a common ovoid
sized to the broader 1P-OPM FWHM for both modalities, isolating background
rejection from the resolution difference and giving a conservative estimate
for 2P-OPM, which integrates additional background around its tighter spot.
SNR was computed for $N = 20$ matched puncta from the same embryo,
normalized to the mean 1P-OPM SNR, and compared with a two-sided Wilcoxon
signed-rank test (Supplementary Fig.~\ref{fig:snr}).

\subsubsection*{Transcription-site spot tracking and signal integration}
Transcription spot segmentation, tracking, and quantification from \textit{in vivo} data were performed in Imaris (Oxford Instruments) using spot segmentation. Spot detection used background subtraction with estimated diameters of 1~$\si{\micro\meter}$ in $xy$ and 1.25~$\si{\micro\meter}$ in $z$. Candidate spots were filtered by quality and intensity sum ($>\!2230$) to retain low-SNR transcription puncta while suppressing isolated noise.

Tracks were computed with the Brownian motion model (max distance \(=\) \(2.36~\si{\micro\meter}\), max gap size \(= 3\); no gap filling). Short tracks were excluded using Filter Tracks: Track Duration.

For each accepted track and time point, signal was quantified as the intensity sum inside a fixed, ellipsoidal ROI
matching the detection diameters.
Local background was the mean intensity in a concentric shell with outer diameters $2\times$ the ROI (shell = outer minus inner).
The resulting background-subtracted traces were taken directly (no temporal gap filling).

\subsubsection*{Single-molecule detection and tracking}
Single molecules from \textit{in vivo} data were detected and tracked in TrackMate~7.12.2 (Fiji)~\cite{Ershov_2022,Schindelin_2012}. Detection used the Difference-of-Gaussian (DoG; LoG approximation) detector with an estimated object diameter = 0.347~\si{\micro\meter}, quality threshold = 3, sub-pixel localization enabled, and no median pre-filtering.

Tracks were built with a linking max distance = 1~\si{\micro\meter}, gap-closing max distance = 1~\si{\micro\meter}, and gap-closing max frame gap = 1 (simple LAP tracker; non-branching; gap closing only).

\subsubsection*{Photon calibration and localization precision}
Spot intensities (digital numbers, DN) from TrackMate were converted to detected photons using the camera’s conversion gain $g$ ($e^{-}\!/\mathrm{DN}$) and quantum efficiency $\eta$.
We measured an electronic offset $O$ (DN) by averaging over 100 dark images acquired with the same exposure and no illumination.
For each localization, the detected signal photons and the residual background per pixel (photons/pixel) are
\begin{equation}\label{eq:calib}
\begin{aligned}
N_\gamma &= \frac{g}{\eta}\Big[(I_{\mathrm{spot}}-O) - (I_{\mathrm{bg}}-O)\Big],\\
b_\gamma &= \frac{g}{\eta}\,\big(I_{\mathrm{bg/pix}}-O\big),
\end{aligned}
\end{equation}
where $I_{\mathrm{spot}}$ is the integrated spot intensity (DN) from the DoG fit, and $I_{\mathrm{bg}}$ and $I_{\mathrm{bg/pix}}$ are the local background (DN) estimated around the spot (total within the spot ROI and per pixel, respectively; $I_{\mathrm{bg}} = N_{\mathrm{mask}}\, I_{\mathrm{bg/pix}}$, where $N_{\mathrm{mask}}$ is the number of pixels in the spot integration mask).

Localization precision was estimated using the Cramér–Rao lower bound for a 2D Gaussian PSF following Thompson \textit{et al}.~\cite{Thompson_2002} and the closed-form refinements of Mortensen \textit{et al}.~\cite{Mortensen_2010}.
With pixel size $a$ (in $\mu$m), PSF standard deviations in camera coordinates $(x,y')$ denoted $\sigma_{\mathrm{PSF}x}$ and $\sigma_{\mathrm{PSF}y'}$ (in $\mu$m), detected photons per localization $N_\gamma$, and residual background per pixel $b_\gamma$ (photons/pixel), the per-localization variances are
\begin{equation}
\label{eq:crlb}
\begin{aligned}
\sigma_{\mathrm{loc},x}^{2}
&= \frac{\sigma_{\mathrm{PSF}x}^{2}+a^{2}/12}{N_\gamma}
 + \frac{8\pi\,\sigma_{\mathrm{PSF}x}^{4}\,b_\gamma}{a^{2}N_\gamma^{2}},\\
\sigma_{\mathrm{loc},y'}^{2}
&= \frac{\sigma_{\mathrm{PSF}y'}^{2}+a^{2}/12}{N_\gamma}
 + \frac{8\pi\,\sigma_{\mathrm{PSF}y'}^{4}\,b_\gamma}{a^{2}N_\gamma^{2}}.
\end{aligned}
\end{equation}
PSF widths were obtained from smFISH puncta measured in the fly embryo (Fig.~\ref{fig:main4}), to recapitulate the scattering and aberrations present at depth in tissue.
Because the detection plane is tilted by $\theta=45^\circ$ in the $yz$-plane (i.e., about $x$; see Fig.~\ref{fig:main1}c and Supplementary Fig.~\ref{fig:2P-OPM_geo-theory-NA}), the $y'$ PSF mixes  high-resolution, in-plane lateral blur with broader axial blur, degrading localization precision along $y'$ relative to $x$ for otherwise identical photon/background conditions.
To account for this, we used the value
$\sigma_{\mathrm{PSF}y'}=\sqrt{\sigma_y^{2}\cos^{2}\theta + \sigma_z^{2}\sin^{2}\theta}$,
while $\sigma_{\mathrm{PSF}x}$ is simply the $x$ width.

Using Eq.~\eqref{eq:crlb}, we computed a per-spot localization precision from each spot’s own $N_\gamma$ and $b_\gamma$ and then averaged across all localizations:
\begin{equation}
\label{eq:meanprec}
\begin{aligned}
\sigma_{\mathit{x}}   &= \big\langle \sigma_{\mathrm{loc},x} \big\rangle,\\
\sigma_{\mathit{y}'} &= \big\langle \sigma_{\mathrm{loc},y'} \big\rangle.
\end{aligned}
\end{equation}

\subsubsection*{Ensemble diffusion coefficient}
Tracks were analyzed at a time interval $\Delta t=50$~ms (20~Hz).
For each consecutive time step we computed $(\Delta x,\Delta y')$ and the squared displacement $\Delta r^{2}=(\Delta x)^2+(\Delta y')^2$.
Averaging over all consecutive steps gives
\begin{equation}
\label{eq:msd1}
\begin{aligned}
\mathrm{MSD}_{\Delta t} &= \big\langle \Delta r^{2} \big\rangle,\\
\mathrm{MSD}_{\Delta t}^{\mathrm{corr}} &= \mathrm{MSD}_{\Delta t} - 2\big(\sigma_{\mathit{x}}^{2}+\sigma_{\mathit{y}'}^{2}\big),
\end{aligned}
\end{equation}
which removes localization variance along $x$ and $y'$.
Assuming 2D free diffusion at short intervals,
\begin{equation}
\label{eq:DfromMSD1}
D \;=\; \frac{\mathrm{MSD}_{\Delta t}^{\mathrm{corr}}}{4\,\Delta t}.
\end{equation}
Uncertainty on $\mathrm{MSD}_{\Delta t}$ is the standard error across consecutive-step squared displacements; uncertainty on $D$ follows from Eq.~\eqref{eq:DfromMSD1}.

\subsubsection*{MSD versus time interval}
For Fig.~\ref{fig:main5}h, we extended the analysis to time intervals $\tau=L\,\Delta t$ with $L=1, 2, 3, 4$.
Within each track, positions separated by $\tau$ were paired to compute
\begin{equation}
\label{eq:msdL}
\begin{split}
\mathrm{MSD}(\tau) \;=\;
&\Big\langle \big[x(t{+}\tau)-x(t)\big]^2 \\
&+ \big[y'(t{+}\tau)-y'(t)\big]^2 \Big\rangle
\;-\; 2\big(\sigma_{\mathit{x}}^{2}+\sigma_{\mathit{y}'}^{2}\big),
\end{split}
\end{equation}
averaged over all valid pairs at that $\tau$ across tracks.
Points show mean $\pm$ standard error; we display the first four intervals (50–200~ms) to avoid bias from dwindling statistics at longer $\tau$.

\subsubsection*{Heterogeneity index}
For each boxed $xy$ MIP in Fig.~\ref{fig:main6}b with pixel intensities $I_j$, let $\bar{I}$ and $\sigma_I$ denote the mean and standard deviation as defined above (Eq.~\eqref{eq:mean}–\eqref{eq:contrast}). 
Defining the local dynamic range
\begin{equation}
R = I_{\max} - I_{\min}.
\end{equation}
We quantify clustering by a dimensionless heterogeneity index
\begin{equation}
\label{eq:Hdef}
H = \frac{\sigma_I^2}{R^2}.
\end{equation}

$H$ was computed per $xy$ MIP for stimulated and unstimulated conditions.
\(H\) is invariant to global multiplicative scaling and specifically reflects changes in spatial unevenness (more/larger/brighter puncta) rather than overall brightness.
In contrast, raw variance $\sigma_I^2$ increases with global gain and can overstate clustering when images are uniformly brighter.

\subsubsection*{Visualization and Software}
Figures and videos were made with FIJI 2.16.0 (ImageJ 1.54p) \cite{Schindelin_2012}, 3Dscript 0.2 \cite{Schmid_2019}, Imaris 10.2 (Oxford Instruments), R 4.5.1 \cite{RCoreTeam2025}, and MATLAB R2020a--R2024a (MathWorks) software.

\subsection*{Statistics and Reproducibility}
No statistical method was used to predetermine sample size.
Data exclusions were limited to the bead fit-quality (Fig.~\ref{fig:main2}d) and track-duration (Fig.~\ref{fig:main5}d,f--h) criteria described above; no other data were excluded.
The experiments were not randomized, and the investigators were not blinded to allocation during experiments and outcome assessment.
Signal-to-noise ratios were compared with a two-sided Wilcoxon signed-rank test (Supplementary Fig.~\ref{fig:snr}).
All experiments were repeated independently three times with similar results; the images shown are representative.
A completed Light Microscopy Reporting Table is provided as Supplementary Data~1.

\section*{Data availability}
The opto-mechanical computer-aided design (CAD) model and associated parts, representative image data, and associated videos (Supplementary Movies~\href{https://kdizon.github.io/2P-OPM/movies/WF-embryo_movie.mp4}{1}--\href{https://kdizon.github.io/2P-OPM/movies/SM_movie.mp4}{4}) are available at \href{https://github.com/kdizon/2P-OPM}{github.com/kdizon/2P-OPM}.
Owing to their large size, the remaining datasets are available from the corresponding author.
Source data are provided with this paper.

\section*{Code availability}
Python code for instrument control/acquisition (fastMC) and volume reconstruction are available at \href{https://github.com/kdizon/2P-OPM}{github.com/\allowbreak kdizon/\allowbreak 2P-OPM}.



\bibliography{references}

\section*{Acknowledgements}
We thank Thomas Gregor (T.G.), Po-Ta Chen, Miloš Nikolić, and other members of the Gregor laboratory, where the experiments were performed.
Their measurements, and the limits we met in trying to extend them, motivated this microscope.
We thank Steven Lowe for machining support and Nezih Dural for bonding the window to O2.
We are grateful to Marianne Bauer for a theorist's perspective on precision in gene regulation, and to Joshua Shaevitz for taking a chance before there was a project.

\section*{Funding}
This work was supported in part by the U.S. National Science Foundation, through the Center for the Physics of Biological Function (PHY-1734030) and RECODE (2134935 to J.E.T.); the National Institutes of Health (R01GM097275 and U01DA047730 to T.G.; U01DK127429 to T.G. and J.E.T.; R01GM144362 to J.E.T.); and Princeton University's Dean for Research Innovation Fund for New Ideas in the Natural Sciences (to K.K.-D. and T.G.).
K.K.-D. was supported by the Robert H. Dicke Fellowship in Experimental Physics from Princeton University.
H.M.M. was supported by the Lewis--Sigler Scholars Program from Princeton University and a Burroughs Wellcome Fund Career Award at the Scientific Interface.

\section*{Author Contributions}
Project conception and supervision: K.K.-D.
Microscope design, construction, and validation: K.K.-D. and Y.C.
Instrument control software: K.K.-D. and A.D.
Biological sample preparation, methodology, and investigation: K.K.-D., Y.C., A.D., S.R., P.H., T.B., R.H.T., J.E.T., and H.M.M.
Analysis: K.K.-D.
Data visualization, figure, and movie preparation: K.K.-D., Y.C., and A.D.
Manuscript preparation: K.K.-D., with input from all coauthors.
Funding acquisition: K.K.-D., J.E.T., and H.M.M.

\section*{Competing interests}
Portions of the technology described herein are covered by a U.S. provisional patent application (No. 63/909,357), assigned to Princeton University (inventor: K.K.-D.).
K.K.-D. is founder of Coastal Applied Physics \& Engineering, and holds equity interests in Coastal Applied Physics \& Engineering and Kanvas Biosciences.
Y.C. is an employee at Triplet Imaging and holds equity interests.
J.E.T. is a scientific advisor for Prolific Machines and Nereid Therapeutics.
H.M.M. is a cofounder and scientific advisor for C16 Biosciences.
The remaining authors declare no competing interests.


\clearpage

\onecolumngrid
\clearpage
\section*{Supplementary information}

\renewcommand{\figurename}{Supplementary Figure}
\renewcommand{\tablename}{Supplementary Table}
\renewcommand{\thefigure}{\arabic{figure}}
\renewcommand{\thetable}{\arabic{table}}
\setcounter{figure}{0}
\setcounter{table}{0}
\makeatletter
\renewcommand{\fnum@figure}{\textbf{\figurename~\thefigure}}
\renewcommand{\fnum@table}{\textbf{\tablename~\thetable}}
\makeatother

\begin{figure}[!h]
\includegraphics[scale=0.705]{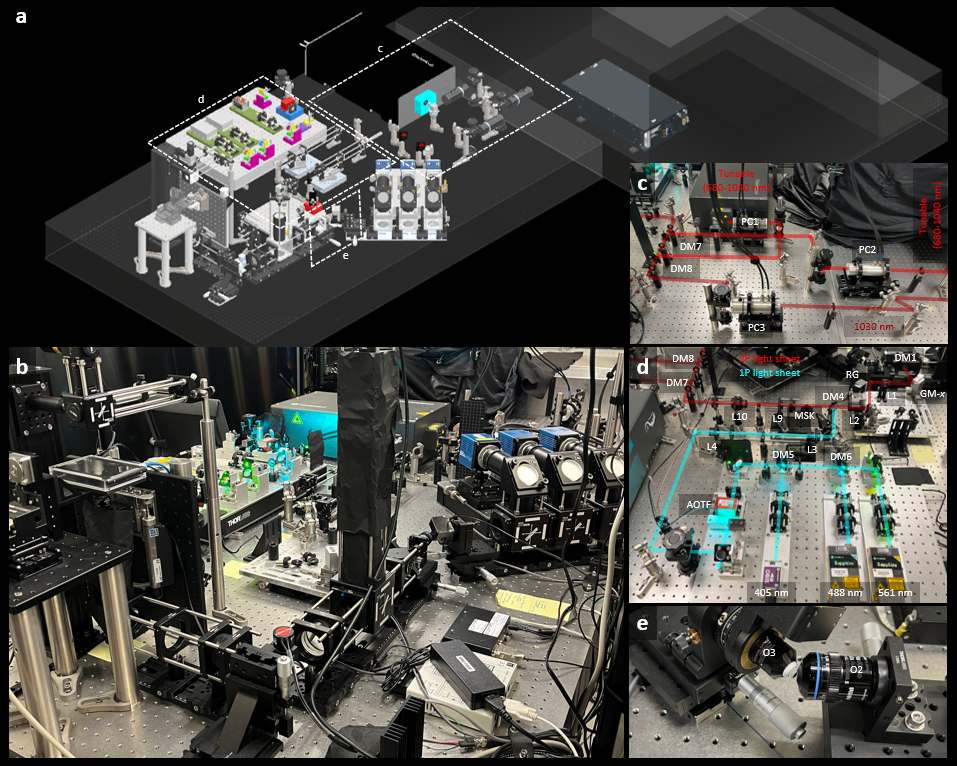}
\caption{\textbf{Opto-mechanical design and implementation of 2P-OPM.}\\
\textbf{(a)} 3D opto-mechanical solid model of 2P-OPM. Complete details of the optical system are given in Methods and Fig.~1a.\\
\textbf{(b)} Photograph of the assembled microscope.\\
\textbf{(c)} Zoomed-in photograph of \textbf{(a)} showing the three Pockels cells used for intensity modulation of each corresponding ultrafast laser, and the broadband and dichroic mirrors used for NIR 2P beam combining.\\
\textbf{(d)} Zoomed-in photograph of the 1P and 2P light-sheet excitation path, from \textbf{(a)}, showing the visible CW lasers, the broadband and dichroic mirrors used for beam combining, the AOTF used to select the 1P wavelengths and to control their amplitudes, as well as the beam conditioning optics used for generating 1P and 2P light sheets.\\
\textbf{(e)} Zoomed-in photograph of the remote-focus subsystem, from \textbf{(a)}, with O2 and O3 oriented at a $45^\circ$ angle between their optical axes. An ultra-thin N-BK7 window (Edmund Optics, 22-043; cut to approximately the diameter of the objective front face) was secured to the mechanical surface of O2 at three contact points with adhesive. The mechanical surface provides an approximately flat mounting reference, sufficient to position the window perpendicular to the optical axis to within the tolerances required for near-diffraction-limited performance. O2's correction collar was set to 0.2 mm to match the window thickness.}
\label{fig:2P-OPM_CAD+photo}
\end{figure}
\clearpage

\begin{figure*}
\includegraphics[scale=0.565]{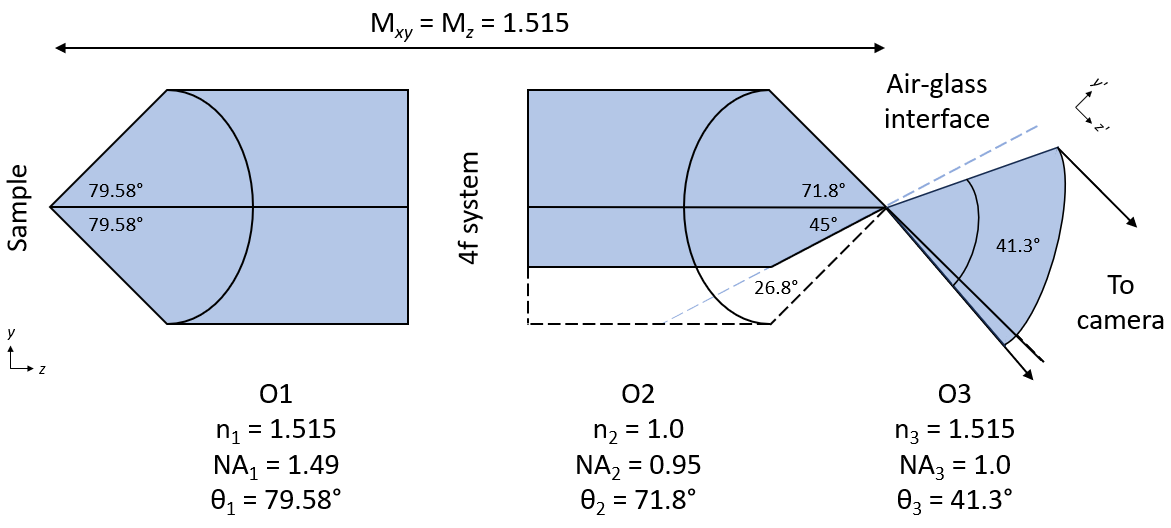}%
\caption{\textbf{Geometric illustration of the theoretical effective detection NA along the $y$’ axis.}\\
Light cones collected by each objective are indicated in blue; black lines represent the paraxial and marginal light rays.
The intermediate image at the focal space of O2 is magnified by NA$\textsubscript{1}$/NA$\textsubscript{2}$ = 1.515× both laterally and axially to minimize aberrations.
Nearly all of the light transmitted through O2 is refracted at the air-glass interface (blue dashed line) and enters O3, except for a small portion of light that is clipped (zone between the blue and black dashed lines).
Along the $x$-axis, i.e., normal to the tilt direction, the theoretical effective detection NA is estimated as $1.515 \cdot \sin(71.8^\circ)$, so that 1.44 of O1's nominal NA of 1.49 can be used.
Along the tilted $y$-axis, light is clipped at the air-glass interface; the upper bound for the theoretical effective NA in $y$ is estimated as $1.515 \cdot \sin( (71.8^\circ + 45^\circ)/2 )$ = 1.29.\\
A larger tilt is advantageous for light-sheet coverage: more NA is available to thin the sheet, and more of its usable length projects onto the axial direction, raising the imaging-volume height reachable per sweep~\cite{Sapoznik}. Shallower angles improve emission collection but constrain the minimum light-sheet thickness, reducing sectioning. We use $45^\circ$, the maximum acceptance angle of O3, to maximize $z$-depth coverage for axially extended specimens, accepting the reduced emission collection that a larger tilt brings; $45^\circ$ also simplifies alignment and reconstruction.}
\label{fig:2P-OPM_geo-theory-NA}
\end{figure*}
\clearpage

\begin{figure*}
\includegraphics[scale=0.9]{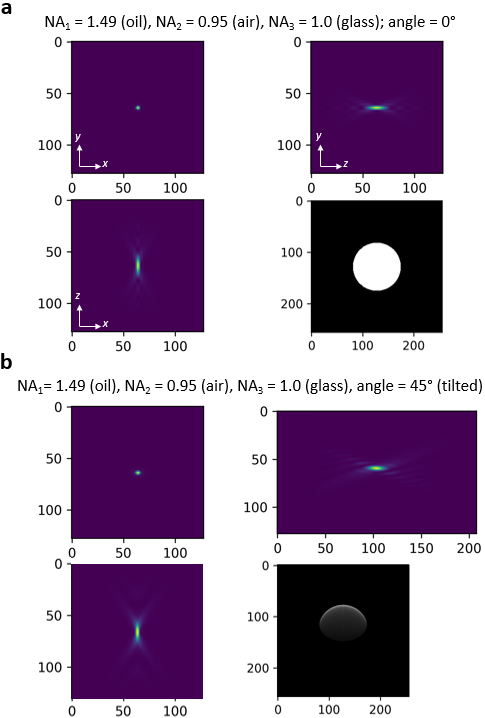}%
\caption{\textbf{Simulations of 2P-OPM detection point spread and pupil functions.}\\
\textbf{(a)} Simulated $xy$ (top left), $xz$ (bottom left), and $yz$ (top right) cross sections of the PSF and pupil function (bottom right) for the high-NA 2P-OPM in straight transmission. The $xyz$ FWHM values of the PSF:  $181.8 \times 181.8 \times 872.3$ nm$^{3}$.\\
\textbf{(b)} Simulated $xy$ (top left), $xz$ (bottom left), and $yz$ (top right) cross sections of the PSF and pupil function (bottom right) for the high-NA 2P-OPM in the tilted configuration.
The $xyz$ FWHM values of the PSF with a 45$^\circ$-tilt: $179.9 \times 237.3 \times942.5$~nm$^{3}$.
The FWHM values along $x$ are comparable between the straight and the tilted configurations, suggesting that the full angular aperture of O1 and O2 are utilized.
The FWHM values along the other two directions ($y$ and $z$) are slightly wider in the tilted configuration, due to the asymmetry of the pupil function, and hence the effective resolution is slightly worse.}
\label{fig:2P-OPM_sim}
\end{figure*}
\clearpage

\begin{figure*}
\includegraphics[scale=0.75]{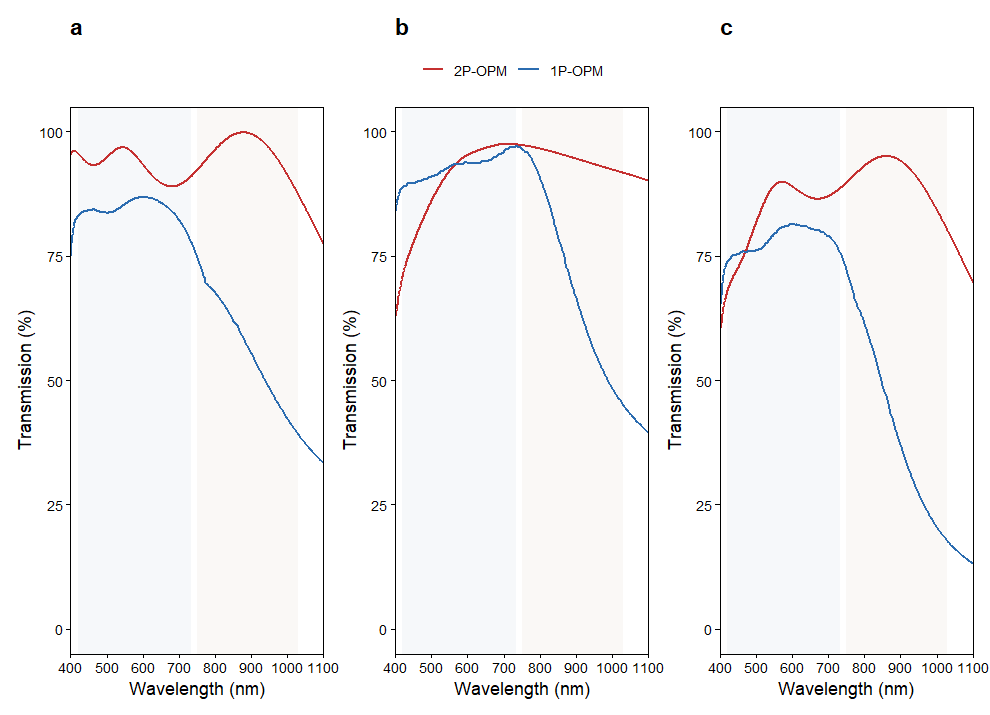}%
\caption{\textbf{Relay-lens transmission across wavelength.}\\
Transmission of the broadband VIS--NIR relay optics used here versus the visible-optimized optics of the prior high-NA OPM \cite{Sapoznik}, decomposed into \textbf{(a)}~scan lenses, \textbf{(b)}~tube lenses, and
\textbf{(c)}~the combined shared-path relay (scan $\times$ tube). Components are listed in Table~1; each curve is the product of its constituent lens transmissions, computed from manufacturer transmission or AR-coating reflectance data. No transmission curve is published for the prior LSM03-VIS scan lens, so its contribution to \textbf{(a)} and \textbf{(c)} is estimated from reflectance (neglecting bulk absorption and scatter) and held flat beyond its 800~nm characterization limit; both choices favor \cite{Sapoznik}, so its scan-lens and combined curves are upper bounds. Shaded bands mark the fluorescence emission collection range (420--734~nm) and the 2P excitation range (750--1030~nm).\\
Source data are provided as a Source Data file.}
\label{fig:transmission}
\end{figure*}
\clearpage

\begin{figure*}
\includegraphics[scale=0.9]{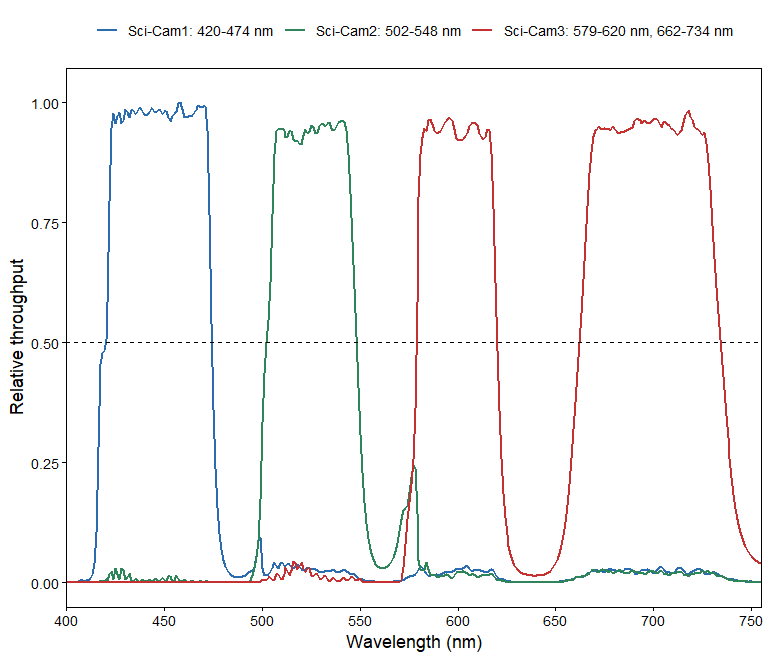}%
\caption{\textbf{Emission channels for 2P-OPM.}\\
Relative throughput to each sCMOS camera (Sci-Cam1--3), computed from manufacturer dichroic spectra by multiplying transmissions along the dichroic-mirror tree (reflections taken as $R{=}1{-}T$). Curves share a single global normalization (unitless); the dashed line marks the 50\% threshold used to define the passbands shown in the legend. See Table~1 for parts list and Fig.~1a for optical diagram.\\
Source data are provided as a Source Data file.}
\label{fig:spectra}
\end{figure*}
\clearpage

\begin{figure*}
\includegraphics[scale=0.975]{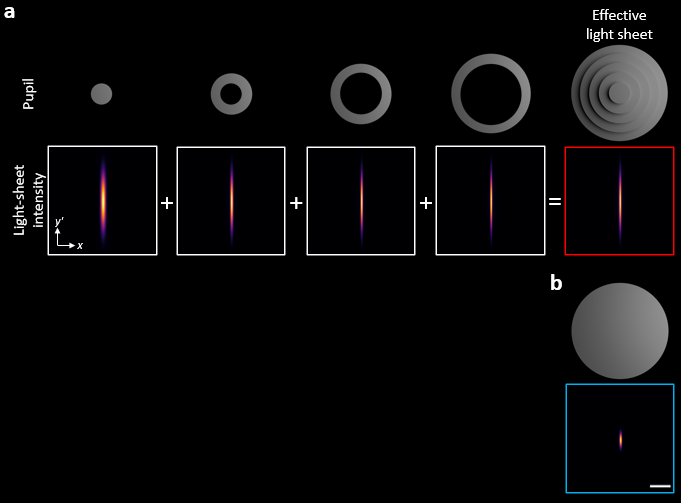}%
\caption{\textbf{Principle and simulations of Bessel-like light sheet.}\\
\textbf{(a)} A series of apertures at the pupil (top row), each of different diameter, produce the same axial extent (second row). The (left column) aperture produces a Gaussian-like focus, whereas the annular apertures (middle columns) produce a Bessel-like focus.
Inspired by the pupil plane approach to extended focus from Gustafsson \textit{et al}. \cite{Abrahamsson_2006}, the layer-cake phase mask (right column) segments the pupil into multiple zones.
The phase mask consists of multiple concentric glass disks, each $\sim\!\!300$ $\si{\micro\meter}$ thick, creating a time delay between zones longer than the pulse duration of the ultrafast (femtosecond pulse) laser, ensuring negligible temporal overlap and preventing interference between pulses \cite{ChenChakraborty_2020,ExD_KKD}.
This results in independent beamlets that add incoherently to produce an axially elongated Bessel-like beam, with minimal broadening to the thickness of the beam focus (right column).
Because the superposition is incoherent and the power is spread along this longer beam, exciting the same 2P signal as a tightly focused Gaussian beam requires correspondingly higher average power.
The amount of DOF scales linearly to the number of layers in the mask (in this case $\sim\!4\times$), and can thus be tuned to the desired light-sheet thickness and length.\\
\textbf{(b)} A conventional high-NA Gaussian beam light sheet, where different wavevectors interfere and add coherently to produce a tight focus, for comparison. 
Scale bar, 2 $\si{\micro\meter}$.}
\label{fig:layer-cake_sim}
\end{figure*}
\clearpage

\begin{figure*}
\includegraphics[scale=0.675]{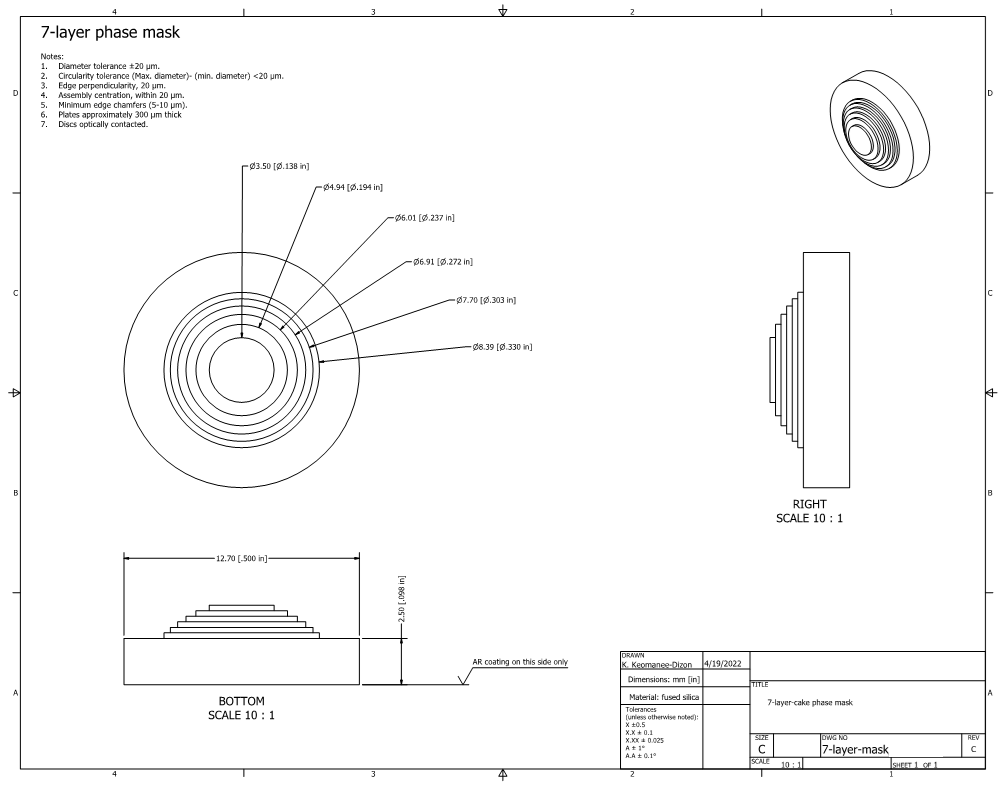}%
\caption{\textbf{Layer-cake phase mask design and drawing.}\\
In the experiments presented, we under-filled the phase mask and used only 4 of the innermost layers, extending the depth of focus accordingly.}
\label{fig:layer-cake_design}
\end{figure*}

\begin{figure*}
\includegraphics[scale=0.622]{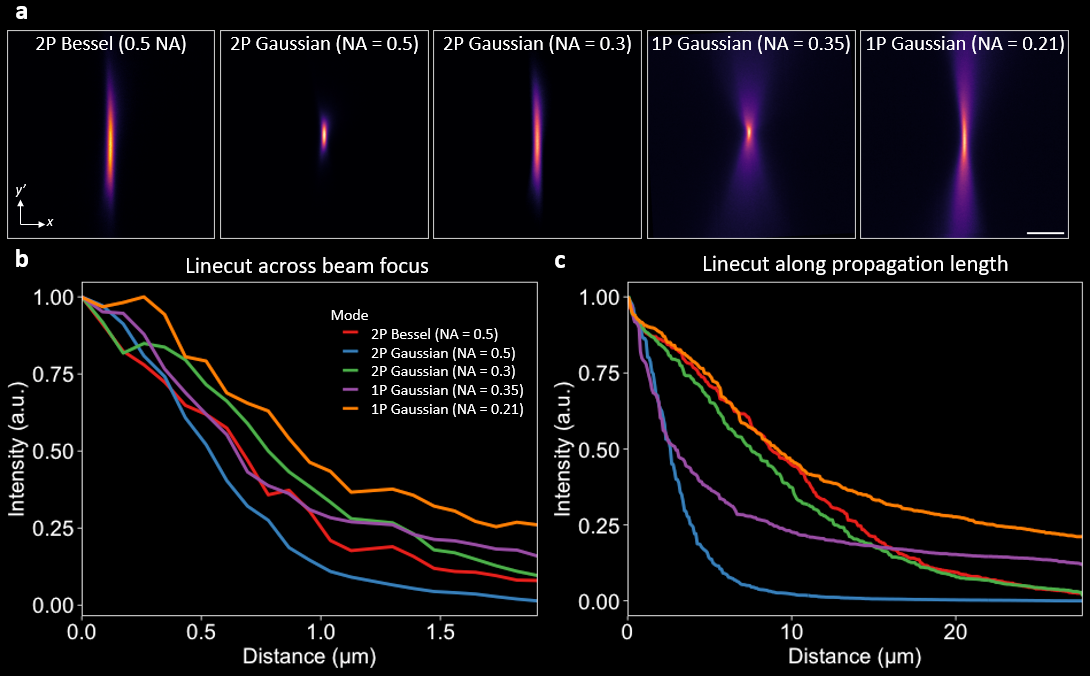}%
\caption{\textbf{Comparison and characterization of light-sheet profiles.}\\
\textbf{(a)} Experimental images of fluorescence excited by 2P Bessel (column 1), 2P high-NA Gaussian (column 2), 2P low-NA Gaussian (column 3), 1P high-NA Gaussian (column 4), and 1P low-NA Gaussian (column 5) focused beams, which are rapidly scanned in the $x$ direction to create light sheets. Scale bar, 2 \si{\micro\meter}.\\
\textbf{(b)} Intensity profiles of \textbf{(a)} at the beam focus. The FWHM values are 2P Bessel: $\sim\!\!0.6$ $\si{\micro\meter}$; 2P Gaussian high NA: $\sim\!\!0.58$ $\si{\micro\meter}$; 2P Gaussian low NA: $\sim\!\!0.85$ $\si{\micro\meter}$; 1P Gaussian high NA: $\sim\!\!0.62$ $\si{\micro\meter}$; 1P Gaussian low NA: $\sim\!\!0.95$ $\si{\micro\meter}$. a.u.: arbitrary units.\\
\textbf{(c)} Intensity profiles of \textbf{(a)} along the propagation length. The FWHM values are 2P Bessel: $\sim\!\!10$ $\si{\micro\meter}$; 2P Gaussian high NA: $\sim\!\!3.2$ $\si{\micro\meter}$; 2P Gaussian low NA: $\sim\!\!9.8$ $\si{\micro\meter}$; 1P Gaussian high NA: $\sim\!\!3.4$ $\si{\micro\meter}$; 1P Gaussian low NA: $\sim\!\!10.2$ $\si{\micro\meter}$. a.u.: arbitrary units.}
\label{fig:2P-OPM_beams}
\end{figure*}
\clearpage

\begin{figure*}
\includegraphics[scale=0.735]{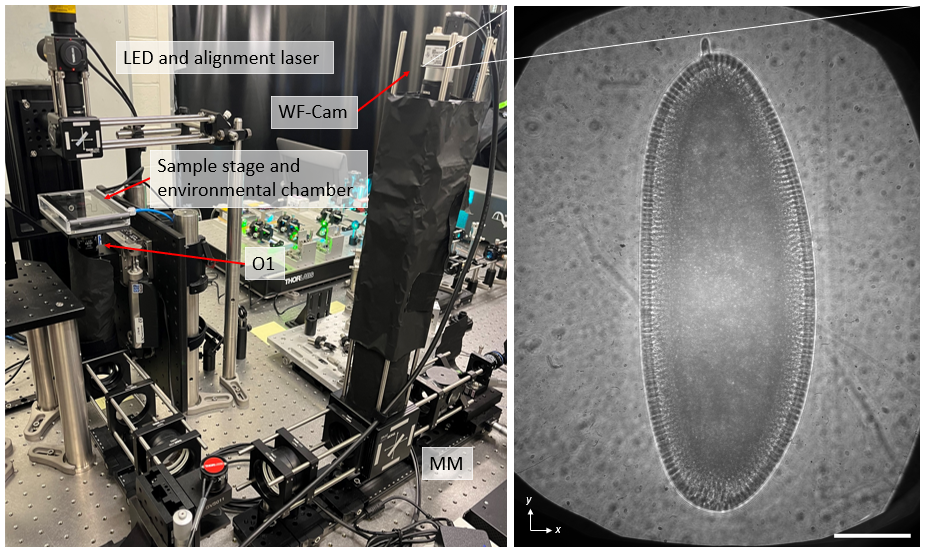}%
\caption{\textbf{Wide-field illumination and detection.}\\
Photograph of the wide-field illumination module and the wide-field detection camera (WF-Cam) of the 2P-OPM.
455 nm and 625 nm LED light, and/or 405 nm, 488 nm, 561 nm laser light, are used for trans-illumination of the sample (e.g., for photoactivation) and/or alignment of the optical system. A movable mirror (MM) is used to direct light to WF-Cam for specimen view-finding and inspection, recording macroscopic behavior, as well as a photoperturbation indicator. 
With an effective magnification of 10×, the WF-Cam yields a $> 530 \times 710~\si{\micro\meter}^{2}$ field of view.
The inset shows an image of a fruit fly embryo, from Supplementary Movie~1. Scale bar, 100 \si{\micro\meter}.}
\label{fig:WF-Cam_photo+embryo}
\end{figure*}
\clearpage

\begin{figure*}
\centering
\includegraphics[scale=0.7]{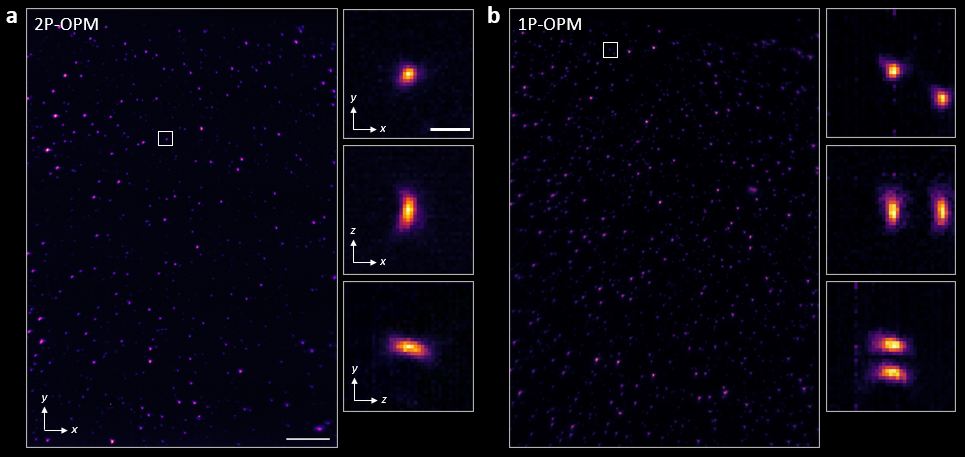}
\caption{\textbf{2P-OPM PSF benchmarking.}\\
\textbf{(a,b)} $xy$ MIPs of a $70 \times 100 \times 15$ \si{\micro\meter}$^{3}$ volume cut from a beads field captured with 2P-OPM \textbf{(a)} and 1P-OPM \textbf{(b)}.  Scale bar, 10 \si{\micro\meter}.\\
Insets show $xy$ (top), $xz$ (middle), and $yz$ (bottom) MIPs of a representative bead from the region indicated by the white box for each respective mode. MIPs use linear contrast, adjusted separately for the fields and insets, so the zoomed views may appear different. Scale bar, 1 \si{\micro\meter}.
Line intensity profiles through these PSFs and the FWHM resolution across the field of view are shown in Fig.~2.
}
\label{fig:2P+1P-OPM_PSF}
\end{figure*}
\clearpage

\begin{figure*}
\includegraphics[scale=0.75]{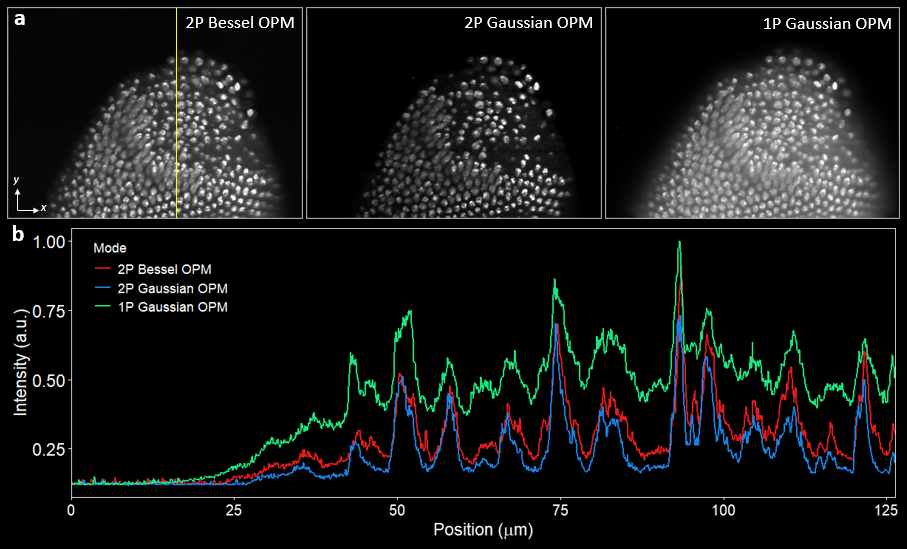}%
\caption{\textbf{Imaging multicellular systems with 2P Bessel, 2P Gaussian, and 1P Gaussian OPM.}\\
\textbf{(a)} $xy$ MIP of a DAPI-stained fruit fly embryo captured with 2P Bessel (left), 2P Gaussian (middle), and 1P Gaussian OPM.\\
\textbf{(b)} Fluorescence intensity profiles along the yellow line shown in \textbf{(a)}. 2P Bessel and 2P Gaussian OPM show progressively improved SNR compared to 1P Gaussian OPM. a.u.: arbitrary units.\\
Source data are provided as a Source Data file.}
\label{fig:2P-OPM_Bessel}
\end{figure*}
\clearpage

\begin{figure*}
\includegraphics[scale=1]{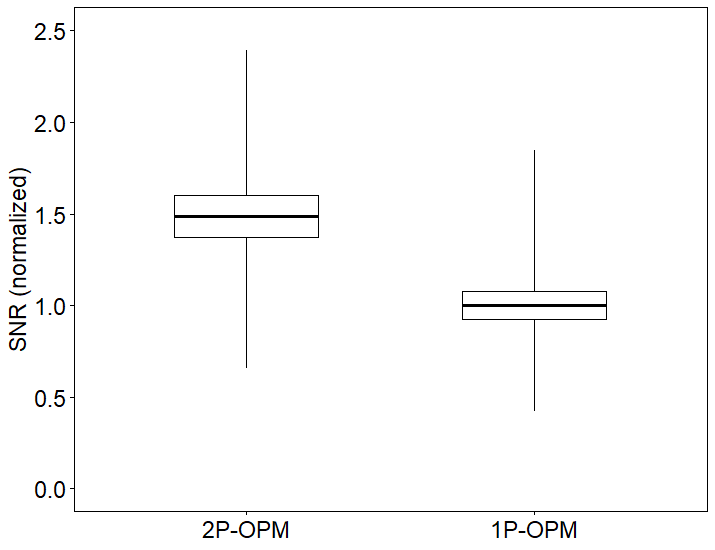}%
\caption{\textbf{smFISH signal-to-noise ratio.}\\
Per-punctum SNR for 2P-OPM and 1P-OPM, computed for $N = 20$
matched puncta from the same embryo and normalized to the mean 1P-OPM SNR.
2P-OPM yields a $\sim\!\!1.5\times$ higher mean SNR, with higher SNR for all 20
puncta (two-sided Wilcoxon signed-rank test, $p = 1.9\times10^{-6}$). The mean absolute SNR is $45.8 \pm 15.9$ for 2P-OPM and $30.8 \pm 10.5$ for 1P-OPM ($N = 20$; mean $\pm$ SD). SNR was integrated over a common ovoid matched to the broader 1P-OPM puncta FWHM (Methods), a conservative choice for 2P-OPM. Box bounds, mean $\pm$ standard error (not percentiles); center value, mean; whiskers, minimum to maximum.\\
Source data are provided as a Source Data file.}
\label{fig:snr}
\end{figure*}
\clearpage

\begin{figure*}
\includegraphics[scale=1]{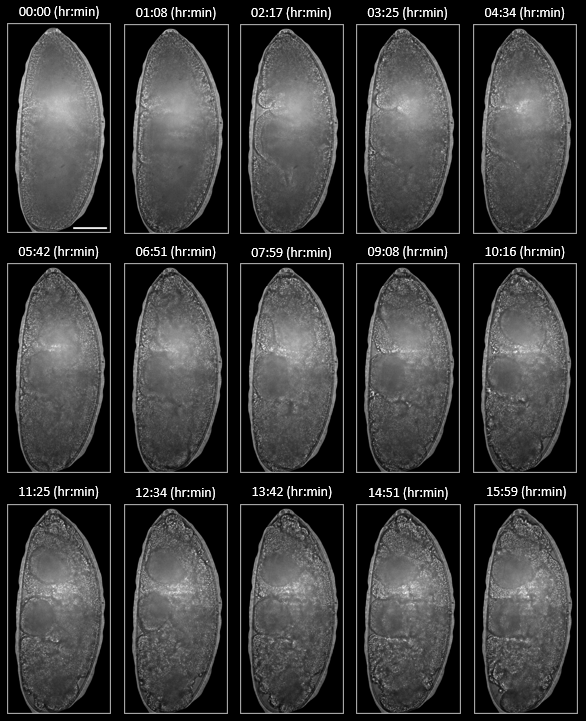}%
\caption{\textbf{16-hr time-lapse of a live \textit{Drosophila} embryo following 2P-OPM imaging.}\\
Bright-field time-series acquired after 2P-OPM imaging to assess sample viability. 16-hr time-lapse consists of 3000 time points at 20 s intervals. The embryo progressed through normal morphogenesis with no phenotypic signs of phototoxicity throughout the recording. Scale bar, 100 \si{\micro\meter}.}
\label{fig:WF-embryo_post2p}
\end{figure*}
\clearpage

\begin{figure*}
\includegraphics[scale=0.7]{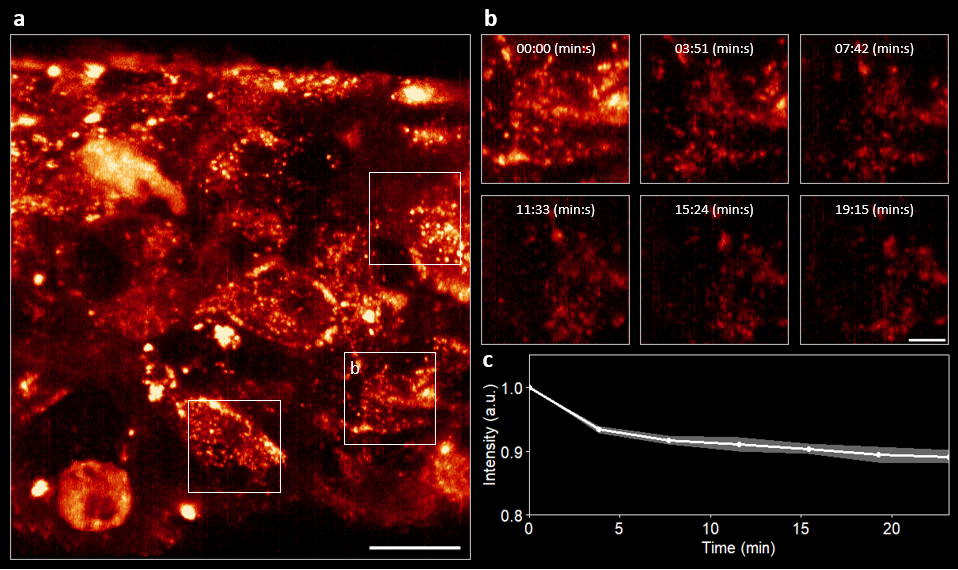}%
\caption{\textbf{2P-OPM imaging at 1030 nm does not activate CRY2-based optogenetic constructs.}\\
\textbf{(a)} $xy$ MIP of FusionRed-OptoEGFR-expressing RPE-1 cells imaged with 1030~nm 2P-OPM in the absence of blue-light stimulation. White boxes indicate the regions shown and analyzed in \textbf{(b)} and \textbf{(c)}. Scale bar, 20~$\mu$m.\\
\textbf{(b)} Zoomed time series of the boxed region \textbf{b} in \textbf{(a)} under continuous 1030~nm illumination (27~min, no blue-light stimulation), showing decreasing fluorescence with no emergence of clustering. Scale bar, 5~$\mu$m.\\
\textbf{(c)} Normalized integrated fluorescence intensity over time for the three regions in \textbf{(a)} (mean $\pm$ standard error, $N=3$ regions), showing a modest monotonic decrease of $\sim$10.9 $\pm$ 1.1\%. No signal amplification or clustering was observed, confirming that 1030~nm excitation does not crosstalk with CRY2-based optogenetic constructs activated at $\sim$850~nm and $\sim$950~nm \cite{Izquierdo_2018,Taslimi_2014}. a.u.: arbitrary units.\\
Source data are provided as a Source Data file.}
\label{fig:opto-control}
\end{figure*}
\clearpage

\begin{figure*}
\includegraphics[scale=0.881]{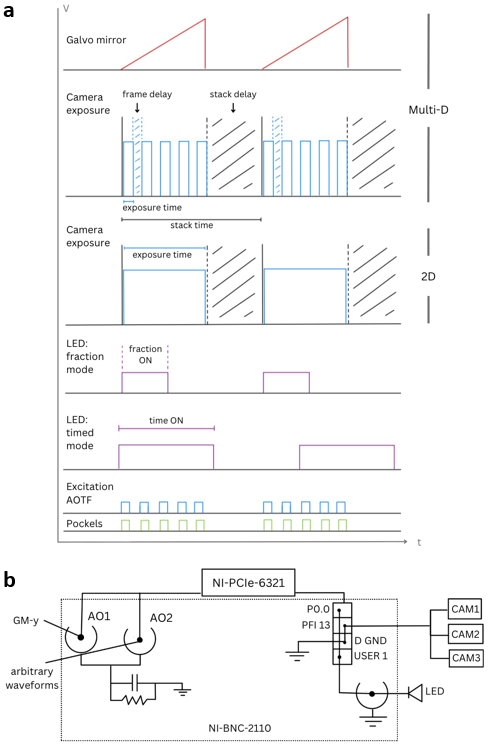}%
\caption{\textbf{Schematic of fastMC control signal sequences and DAQ-device connections.}\\
\textbf{(a)} Top: GM-$y$ is driven by a sawtooth; the ramp advances planes within a stack and the reset occurs during the stack delay (hatched).\\
Middle: Camera exposure waveforms (blue). In 3D (Multi-D) mode, a burst of exposure pulses forms a $y$-stack, separated by a stack delay for galvo retrace; in 2D mode, single exposures repeat with a per-frame delay. Each TTL pulse defines both the start and the duration of the exposure; hatched regions indicate readout/idle time.\\
Lower: Wide-field LED illumination/optogenetic control modes. Fraction mode turns the LED on for a user-defined fraction of each stack; timed mode applies one contiguous on window starting at acquisition.\\
Bottom traces: The AOTF/Pockels are phase-locked to the exposure pulses; excitation is enabled only during camera exposure and blanked during readout and galvo retrace.\\
\textbf{(b)} All timing signals in \textbf{(a)} are generated by the NI X-Series DAQ under fastMC (see Methods) and wired via the NI-BNC-2110.
Analog output ports (AO1, AO2) trigger GM-$y$ to start scanning as defined by the precomputed waveform driven from AO2 (e.g., a signal generator). Digital output ports (PO.0, PFI~13) send concurrent TTL pulses to Sci-Cam1--3 and the LED. Additional cameras can be added by fanning out the same TTL (provided the acquisition PC can sustain the aggregate throughput).}
\label{fig:2P-OPM_timing}
\end{figure*}
\clearpage

\begin{figure*}
\includegraphics[scale=0.57]{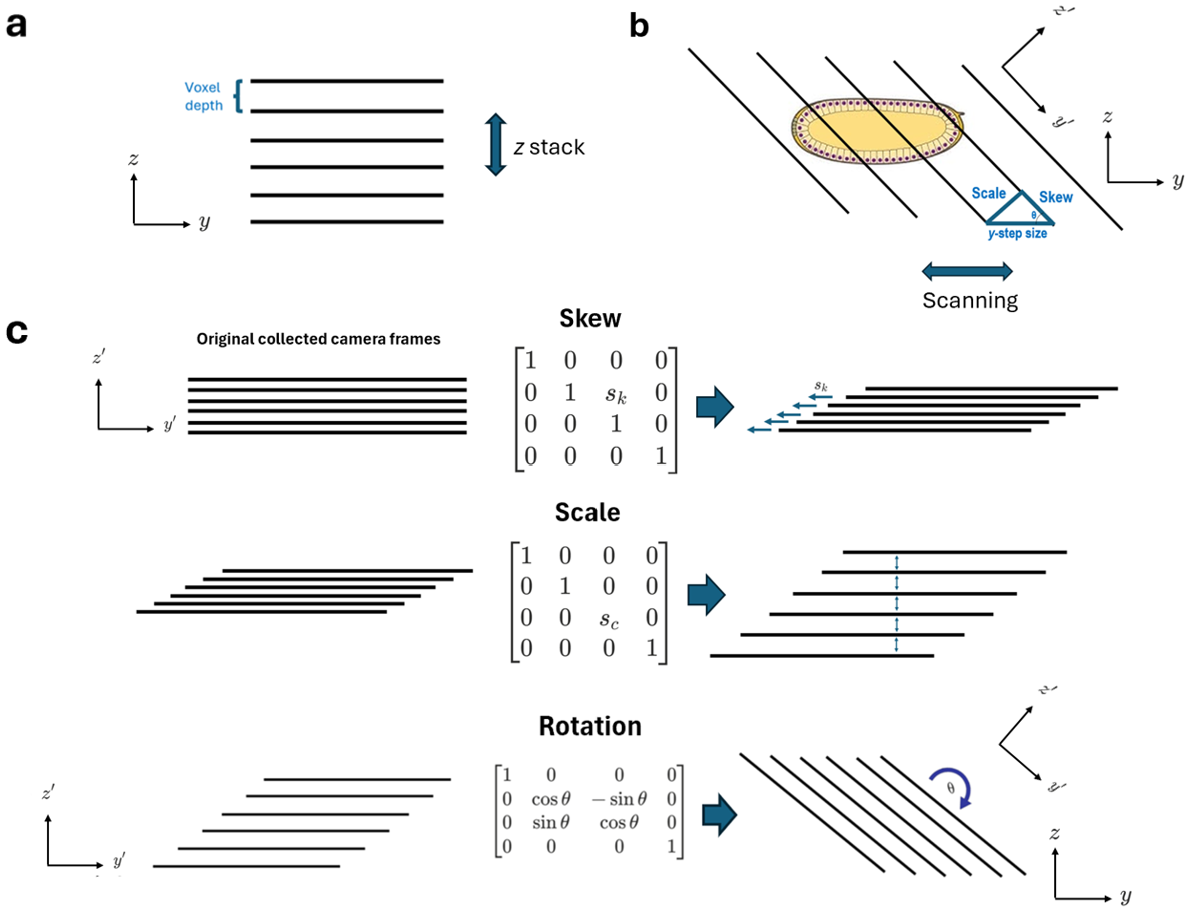}
\caption{\label{fig:recon}\textbf{OPM geometry and reconstruction.}\\
\textbf{(a)} Conventional $z$-stacks.\\
\textbf{(b)} OPM “$y$-stacks”: tilt by $\theta$ produces per-frame offsets
$d_{\mathrm{skew}}=y_{\mathrm{step}}\cos\theta$ (along $y'$) and
$d_{\mathrm{scale}}=y_{\mathrm{step}}\sin\theta$ (along $z'$); $x$ and $y'$ are the in-plane camera axes. See also Fig.~1b-d.\\
\textbf{(c)} Affine sequence Skew ($K$) $\rightarrow$ Scale ($S$) $\rightarrow$ Rotation ($R$)
as in Eqs.~(4)--(9) of the main text, followed by interpolation onto a Cartesian grid (Methods).}
\end{figure*}
\clearpage

\newcommand{\colspace}{\hspace{0.8em}} 
\begingroup
\setlength{\tabcolsep}{6pt}            

\begin{turnpage}
\begin{table*}[p]
  \centering
  \caption{\textbf{Imaging conditions.}}
  \label{tab:imaging}
  \small
  \resizebox{\textwidth}{!}{%
  \begin{tabular}{@{}%
      l @{\colspace} c @{\colspace} l @{\colspace} l @{\colspace} c
      @{\colspace} c @{\colspace} c @{\colspace} c @{\colspace} c
      @{\colspace} c @{\colspace} c @{\colspace} c @{}}
  \toprule
  \textbf{Figure and Video} &
  \textbf{Mode} &
  \textbf{Sample} &
  \textbf{Fluorescent label} &
  \textbf{Temperature} &
  \makecell{\textbf{Excitation}\\$\lambda$ (nm), NA, Power\textsuperscript{†} (mW)} &
  \makecell{\textbf{Voxel}\\$a_x,a_y,a_z$ (nm)} &
  \makecell{\textbf{\textit{y} step (nm)}} &
  \makecell{\textbf{Volume}\\$x,y,z$ (\si{\micro\meter})} &
  \makecell{\textbf{Exposure}\\(ms)} &
  \makecell{\textbf{Emission channel}} &
  \makecell{$\Delta t \times N\textsuperscript{\P}$\\(Time interval (s) $\times$ Time points)} \\
  \midrule
  Fig.~2a,b1--b5   & 2P-OPM & 50 nm beads & Dragon Green & Room & 925, 0.3, 590 & 87, 87, 87 & 87 & 100, 100, 10 & 300 & Sci-Cam2 & 346.2 $\times$ 1 \\
  Fig.~2c,d and Supplementary Fig.~10a   & 2P-OPM & 50 nm beads & Dragon Green & Room & 925, 0.3, 464 & 87, 87, 87 & 87 & 100, 100, 15 & 200 & Sci-Cam2 & 577 $\times$ 1 \\
  Fig.~2c,d and Supplementary Fig.~10b   & 1P-OPM & 50 nm beads & Dragon Green & Room & 488, 0.21, 0.2 & 87, 87, 87 & 87 & 100, 100, 15 & 100 & Sci-Cam2 & 86.55 $\times$ 1 \\
  Fig.~3a  & Wide-field & mESC gastruloid & N/A & Room & 625, WF\textsuperscript{\S}, 45\textsuperscript{‡} & 345, 345 & N/A & 530, 710 & 1500 & WF-Cam & 1.5 $\times$ 1 \\
  Fig.~3a--d and Supplementary Movie~2  & 2P-OPM & mESC gastruloid & DAPI & Room & 750, 0.3, 128 & 87, 87, 87 & 347 & 100, 115, 15 & 300 & Sci-Cam1 & 86.7 $\times$ 1 \\
  Fig.~3a--d and Supplementary Movie~2  & 1P-OPM & mESC gastruloid & DAPI & Room & 405, 0.21, 0.14 & 87, 87, 87 & 347 & 100, 115, 15 & 300 & Sci-Cam1 & 86.7 $\times$ 1 \\
  Fig.~3e  & Wide-field & mESC gastruloid & N/A & Room & 625, WF\textsuperscript{\S}, 45\textsuperscript{‡} & 345, 345 & N/A & 530, 710 & 1500 & WF-Cam & 1.5 $\times$ 1 \\
  Fig.~3e--h  & 2P-OPM & mESC gastruloid &  FOXC1-Alexa Fluor 546 & Room & 1030, 0.18, 33 & 87, 87, 87 & 347 & 71, 130, 30 & 500 & Sci-Cam3 & 144.5 $\times$ 1 \\
  Fig.~3e--h  & 1P-OPM & mESC gastruloid &  FOXC1-Alexa Fluor 546 & Room & 561, 0.12, 0.08 & 87, 87, 87 & 347 & 71, 130, 30 & 500 & Sci-Cam3 & 144.5 $\times$ 1 \\
  Fig.~4a--c & 2P-OPM & Fly, NC 14 & \textit{eve}-ATTO 565 & Room & 1030, 0.3, 38 & 87, 87, 87 & 173 & 100, 115, 15 & 500 & Sci-Cam3 & 289.5 $\times$ 1 \\
  Fig.~4a--c & 1P-OPM & Fly, NC 14 & \textit{eve}-ATTO 565 & Room & 561, 0.21, 0.08 & 87, 87, 87 & 173 & 100, 115, 15 & 500 & Sci-Cam3 & 289.5 $\times$ 1 \\
  Fig.~5a  & Wide-field & Fly, NC 14 & N/A & Room & 625, WF\textsuperscript{\S}, 45\textsuperscript{‡} & 345, 345 & N/A & 530, 710 & 1500 & WF-Cam & 1.5 $\times$ 1 \\
  \multirow{2}{*}{Fig.~5a} & \multirow{2}{*}{2P-OPM} & \multirow{2}{*}{Fly, NC 14}
 & Bcd-eGFP & \multirow{2}{*}{Room}
 & 920, 0.3, 200 & \multirow{2}{*}{173, 173, 173}
 & \multirow{2}{*}{347} & \multirow{2}{*}{100, 115, 15}
 & \multirow{2}{*}{800} & Sci-Cam2
 & \multirow{2}{*}{154.4 $\times$ 10} \\
&  &  & \textit{hb}-MS2-MCP-mCherry &  & 1030, 0.3, 60 &  &  &  &  & Sci-Cam3 & \\
\addlinespace[2pt]
  Fig.~5b,c and Supplementary Movie~3 & 2P-OPM & Fly, NC 12 & \textit{hb}-MS2-MCP-mNeonGreen & Room & 920, 0.3, 80 & 173, 173, 173 & 520 & 100, 105, 15 & 100 & Sci-Cam2 & 19.3 $\times$ 145 \\
  Fig.~5d & 2P-OPM & Fly, NC 12 & \textit{hb}-MS2-MCP-mNeonGreen & Room & 920, 0.3, 80 & 173, 173, 173 & 520 & 100, 115, 15 & 50 & Sci-Cam2 & 9.65 $\times$ 398 \\
  Fig.~5e--h and Supplementary Movie~4 & 2P-OPM & Fly, NC 14 & \textit{hb}-MS2-MCP-mNeonGreen & Room & 920, 0.3, 122 & 173, 173 & N/A & 100, 20 & 50 & Sci-Cam2 & 0.05 $\times$ 50 \\
  \multirow{2}{*}{Fig.~6a--d} & 2P-OPM & \multirow{2}{*}{RPE-1 OptoEGFR cells}
 & \multirow{2}{*}{FusionRed-OptoEGFR} & \multirow{2}{*}{\SI{37}{\celsius}}
 & 1030, 0.3, 10 & \multirow{2}{*}{87, 87, 87}
 & \multirow{2}{*}{347} & \multirow{2}{*}{50, 65, 15}
 & \multirow{2}{*}{1000} & \multirow{2}{*}{Sci-Cam3}
 & 1440 $\times$ 2 \\
& Photoactivation &  &  &  & 455, WF\textsuperscript{\S}, 120\textsuperscript{‡} &  &  &  &  &  & 90 $\times$ 14 \\
\addlinespace[2pt]
  Supplementary Fig.~9 and Supplementary Movie~1   & Wide-field & Fly, NC14 & N/A & Room & 625, WF\textsuperscript{\S}, 45\textsuperscript{‡} & 345, 345 & N/A & 530, 710 & 1500 & WF-Cam & 20 $\times$ 3850 \\
  Supplementary Fig.~11a,b  & 2P Bessel OPM & Fly, NC 14 & DAPI & Room & 750, 0.5, 106 & 87, 87, 87 & 200 & 180, 130, 15 & 400 & Sci-Cam1 & 356.4 $\times$ 1 \\
  Supplementary Fig.~11a,b  & 2P-OPM & Fly, NC 14 & DAPI & Room & 750, 0.5, 8 & 87, 87, 87 & 200 & 180, 130, 15 & 400 & Sci-Cam1 & 356.4 $\times$ 1 \\
  Supplementary Fig.~11a,b  & 1P-OPM & Fly, NC 14 & DAPI & Room & 405, 0.35, 0.28 & 87, 87, 87 & 200 & 180, 130, 15 & 400 & Sci-Cam1 & 356.4 $\times$ 1 \\
  Supplementary Fig.~13  & Wide-field & Fly, stage 6 & N/A & Room & 625, WF\textsuperscript{\S}, 45\textsuperscript{‡} & 345, 345 & N/A & 530, 710 & 1500 & WF-Cam & 20 $\times$ 3000 \\
  Supplementary Fig.~14  & 2P-OPM & RPE-1 OptoEGFR cells & FusionRed-OptoEGFR & \SI{37}{\celsius} & 1030, 0.3, 10 & 87, 87, 87 & 347 & 100, 115, 15 & 800 & Sci-Cam2 & 231.2 $\times$ 7 \\
  \bottomrule
  \end{tabular}}
\par\smallskip
\noindent
\begin{minipage}{\textwidth}
\small 
\textsuperscript{†}\,Unless noted, powers were measured at the back aperture of O1.\\
\textsuperscript{‡}\,LED/photoactivation powers were measured at the sample.\\
\textsuperscript{\S}\,WF: Transmitted wide-field illumination was $\sim\!10$ mm in diameter at the sample, overfilling O1's field of view.\\
\textsuperscript{\P}\,$\Delta t$: time between consecutive frames (2D) or volumes (3D).
For continuous acquisition, $\Delta t$ = acquisition time per frame/volume; if a pause is added, $\Delta t$ includes that pause.
$N$: frames or volumes.
\end{minipage}
\end{table*}
\end{turnpage}
\endgroup
\clearpage

\end{document}